\documentstyle[prd,aps,floats,psfig]{revtex}
\def\lsim{\mathrel{\lower2.5pt\vbox{\lineskip=0pt\baselineskip=0pt
\hbox{$<$}\hbox{$\sim$}}}}
\def\gsim{\mathrel{\lower2.5pt\vbox{\lineskip=0pt\baselineskip=0pt
\hbox{$>$}\hbox{$\sim$}}}}

\newcommand{\jbb}{\ \ \bar{\! \! \bar J}}
\newcommand{\nn}{{\nonumber}}
\newcommand{\ima}{{\mbox{Im}\,}}
\newcommand{\rea}{{\mbox{Re}\,}}
\newcommand{\be}{\begin{equation}}
\newcommand{\ee}{\end{equation}}

\newcommand{\NP}[1]{Nucl.\ Phys.\ {#1}}
\newcommand{\ZP}[1]{Z.\ Phys.\ {#1}}

\newcommand{\PL}[1]{Phys.\ Lett.\ {#1}}

\newcommand{\PR}[1]{Phys.\ Rev.\ {#1}}
\newcommand{\PRL}[1]{Phys.\ Rev.\ Lett.\ {#1}}

\begin{document}
\draft
\input epsf
\renewcommand{\topfraction}{0.8}
\twocolumn[\hsize\textwidth\columnwidth\hsize\csname
@twocolumnfalse\endcsname

\preprint{hep-ph/01XXXXXX}
\title{Meson-meson scattering within
one loop Chiral Perturbation Theory\\ and its unitarization.}
\author{A. G\'omez Nicola and
 J. R. Pel\'aez}
\address{Departamento de F{\'\i}sica Te{\'o}rica II,
  Universidad Complutense de Madrid, 28040-- Madrid,\ \ Spain}
\date{September, 2001}
\maketitle

\begin{abstract}
We present the complete one-loop calculation of all the two meson
scattering amplitudes within the framework of SU(3)
Chiral Perturbation Theory, which includes pions, kaons and the eta.
In addition, we have unitarized these amplitudes
 with the coupled channel Inverse Amplitude Method, which
ensures  simultaneously the good low energy
properties of Chiral Perturbation Theory and  unitarity.
We show how this method provides a remarkable description
of meson-meson scattering data
up to 1.2 GeV including the scattering lengths and
the generation of seven light resonances, which is consistent with
previous determination of the chiral parameters. Particular attention is paid
to discuss the differences and similarities of this work with previous
 analysis in the literature.

\end{abstract}


\vskip2pc]

\section{Introduction}

In the last twenty years, Chiral Perturbation Theory (ChPT)
\cite{weinberg,chpt1,chpt2} has emerged as a powerful tool to
describe the interactions of the lightest mesons. These particles
are considerably lighter than the rest of the hadrons, which is
nowadays understood as a consequence of the spontaneous breaking
of the $SU(3)_L\times SU(3)_R$ chiral symmetry down to
$SU(3)_{L+R}$ that would be present in QCD if the three lightest
quarks were massless. In such case, the light mesons would
correspond to the massless Goldstone bosons  associated to the
spontaneous chiral symmetry breaking. Of course, quarks are not
massless, but their masses are so small compared to the typical
hadronic scales, O(1 GeV), that their explicit symmetry breaking
effect also translates into a small mass for the lightest mesons,
which become pseudo-Goldstone bosons. Hence, the three pions
correspond to the pseudo-Goldstone bosons of the $SU(2)$
spontaneous breaking that would occur if only the
u and d quarks were massless, which is a remarkably good approximation.
Similarly, the meson octet formed by the pions, the kaons and the
eta can be identified with the eight pseudo-Goldstone bosons
associated to the $SU(3)$ breaking, when the $s$ quark  is also
included.

The low energy interactions of pions, kaons and the eta
can be described in terms of an effective Lagrangian that
follows the $SU(3)_L\times SU(3)_R\rightarrow SU(3)_{L+R}$
spontaneous symmetry breaking pattern. If we do not include any
additional field apart from the pseudo-Goldstone bosons, this
description will only be valid for energies much below the scale
where new states appear. That is, the effective ChPT 
Lagrangian provides just a low
energy description. As a consequence we can organize all the possible terms
that respect the symmetry requirements in a derivative (and mass) expansion.
Therefore, any amplitude is  obtained as a perturbative expansion
in powers of the external momenta and the quark masses. The importance
 of this formalism is that  the theory is renormalizable and  predictive,
 in the following sense: all loop divergences
appearing at a given order in the expansion can be absorbed
by a finite number of counterterms, or low energy constants, that
appear in the Lagrangian at that very same order. Thus, order by
order, the theory is finite and depends on a few parameters
that can be determined  experimentally. Once these parameters
are known, any other calculation at that order becomes a prediction.
Basically,  these are the main ideas underlying
ChPT, which has proven very
successful to describe the  low energy hadron phenomenology
(for reviews see \cite{books}).

Despite the success of this approach,  it is unfortunately
limited to low energies (usually, less than 500 MeV).
That is the reason why, over the last few years, there has been
 a growing interest
in extending the applicability range of the chiral expansion
to higher energies. Of course, this requires the use of non-perturbative
methods to  improve the high energy behavior of ChPT amplitudes.
These methods include the explicit introduction of heavier
resonant states in the Lagrangian \cite{explicitresonances},
resummation of diagrams in a Lippmann-Schwinger or Bethe-Salpeter
approach \cite{LS}, or other  methods that unitarize the amplitudes like the
Inverse Amplitude Method (IAM) \cite{Truong,IAM1}. The latter
has been
 generalized to allow for a coupled channel formalism \cite{IAM2},
yielding  a   successful description of the
meson-meson scattering amplitudes up to 1.2 GeV, and even
generating dynamically seven  light resonances.

In principle, these methods recover at low energies the good
properties of ChPT, since they use part
of the perturbative information.
However, it should be noted that, so far, the full results to one loop
 for all the meson-meson scattering processes were not
available in the literature. At present
only the $\pi\pi\rightarrow\pi\pi$ \cite{Kpi},
$K\pi\rightarrow K\pi$ \cite{Kpi}, $\eta\pi\rightarrow\eta\pi$
\cite{Kpi} and
the two independent $K^+K^-\rightarrow K^+K^-$,
$K^+K^-\rightarrow K^0 \bar{K}^0$
\cite{JAPaco}
amplitudes have been obtained in the $SU(3)$ ChPT framework,
although with different procedures and notations.
As a consequence, the IAM has only been applied rigorously
to the $\pi\pi$, $K \bar K$ final states, whereas for a
complete treatment of the whole low energy meson-meson
scattering additional approximations had to be done \cite{IAM2}.
In particular, the lowest order expansion could not be
recovered complete up to $O(p^4)$ thus spoiling the
scattering lengths and, in addition,
 it was not possible to compare directly
with the low energy parameters of
standard ChPT in dimensional regularization and the $\overline{MS}-1$ scheme.

In this work, we have calculated all the meson-meson scattering
amplitudes at one loop in ChPT. There are three amplitudes
that have never appeared published in the literature:
$K\eta\rightarrow K\eta$, $\eta\eta\rightarrow\eta\eta$ and
$K\pi\rightarrow K\eta$. The other five have been
recalculated independently and all of them are
given together in a unified notation, ensuring exact
perturbative unitarity and also correcting
previous misprints. Then, we have applied the coupled channel IAM
 to describe the whole meson-meson scattering below
1.2 GeV, including low energy data like scattering lengths.
This new calculation allows for a direct comparison
with the standard low energy constants of ChPT and that is why we have
made a considerable effort in estimating the uncertainties in all our results.
which are 
in very good agreement with the present determinations obtained
 from low energy data without unitarization. The main differences of
this work with \cite{IAM2} are that we consider the full one-loop
results for the amplitudes, ensuring their finiteness and scale
independence in dimensional regularization, we take into account the
new processes mentioned above and we are able to describe more
accurately the low energy region. This had already been  achieved
for the $\pi\pi$ $K\bar{K}$ system only in \cite{JAPaco}, but
here we complete this task for the whole meson-meson scattering.

The paper is organized as follows. In section \ref{memescat} we review
the main features concerning the meson-meson scattering
calculations at one-loop in ChPT. The final results for the amplitudes
 have been collected in Appendix B due to their length. The
definition of partial waves and unitarity is discussed in section
\ref{parwa}, and the IAM is presented in section \ref{secIAM}.
In section \ref{secdata},
we review the available data on meson-meson scattering.
In sections \ref{secIAM1} and \ref{secIAMfit} we first
use the IAM  with present determinations of the low
energy constants and next we make a fit to the data
commented in section \ref{secdata}.
Our conclusions are summarized in section \ref{secsummary}. Apart from
the amplitudes in Appendix B, we have also collected some  useful formulae
in Appendix A.

\section{Meson meson scattering at one loop}
\label{memescat}

The lowest order Lagrangian for SU(3) Chiral Perturbation Theory
is:
\begin{equation}
{\cal L}_2 = \frac{f_0^2}{4} \langle \partial_{\mu} U^{\dagger}
\partial^{\mu} U +
M_0 (U + U^{\dagger})\rangle,
\label{Lag2}
\end{equation}
where $f_0$ is the pion decay constant in the SU(3) chiral limit and the
angular brackets  stand for the trace of the
$3 \times 3$ matrices. The matrix $U$ collects
the pseudo-Goldstone boson fields $\pi, K, \eta$ through
 $U (\Phi)=\exp (i \sqrt{2} \Phi / f_0)$, where

\begin{equation}
\Phi (x) \equiv
\left(
\begin{array}{ccc}
\frac{1}{\sqrt{2}} \pi^0 + \frac{1}{\sqrt{6}} \eta & \pi^+ & K^+ \\
\pi^- & - \frac{1}{\sqrt{2}} \pi^0 + \frac{1}{\sqrt{6}} \eta & K^0 \\
K^- & \bar{K}^0 &  - \frac{2}{\sqrt{6}} \eta
\end{array}
\right).
\end{equation}
and $M_0$ is the tree level mass matrix.
 Throughout this paper we will be assuming
the isospin limit, so that  $M_0$ is given by
\begin{equation}
M_0 =
\left(
\begin{array}{ccc}
M^2_{0\,\pi} & 0 & 0 \\
0 & M^2_{0\,\pi} & 0 \\
0 & 0 & 2 M^2_{0\,K} - M^2_{0\,\pi}
\end{array}
\right).
\end{equation}
As a matter of fact, from these definitions, it can be easily seen
that the tree level masses satisfy the Gell-Mann--Okubo relation
\cite{GMO}: $4 M_{0\,K}^2-M_{0\,\pi}^2-3 M_{0 \eta}^2=0$, that
will be very useful for simplifying the amplitudes.

\begin{figure}[thbp]
\begin{center}
\hspace*{-.5cm}
\hbox{\psfig{file=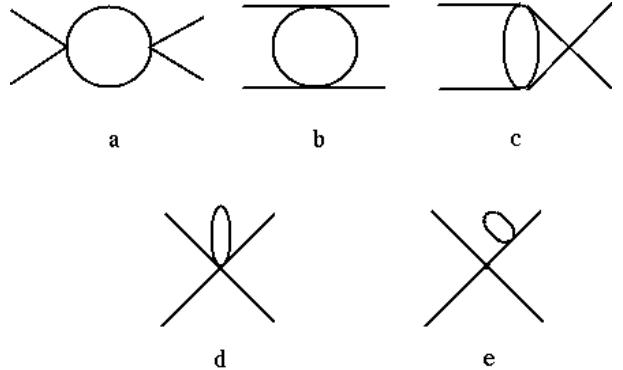,width=8cm}}
\caption{Generic One-loop Feynman diagrams that have to be evaluated
in meson-meson scattering.}
\label{fig1:diagrams}
\end{center}
\end{figure}

From the Lagrangian in Eq.(\ref{Lag2}), one can obtain the
$O(p^2)$ amplitudes just by calculating the corresponding tree
level Feynman diagrams. In order to obtain the $O(p^4)$
contributions, one has to consider loop diagrams, whose generic
topology is given in Fig.\ref{fig1:diagrams}, which will generate
UV divergences. If loop integrals are regularized with
dimensional regularization, which preserves the chiral symmetry
constraints, the divergences can be reabsorbed in the chiral
parameters $L_i$ of the fourth order Lagrangian:
{\setlength{\arraycolsep}{0.01cm}
\begin{eqnarray}
{\cal L}_4&& = L_1 \langle \partial_{\mu} U^{\dagger}
\partial^{\mu} U \rangle^2 +
L_2 \langle \partial_{\mu} U^{\dagger} \partial_{\nu}
U \rangle \langle\partial^{\mu}
U^{\dagger} \partial^{\nu} U \rangle \label{lag4}\\
+&& L_3 \langle \partial_{\mu} U^{\dagger} \partial^{\mu}
U \partial_{\nu} U^{\dagger}
\partial^{\nu} U \rangle +
\! L_4 \langle \partial_{\mu} U^{\dagger} \partial^{\mu} U \rangle
\langle U^{\dagger} M_0\! +\! M_0^{\dagger} U \rangle \nn\\
+&& L_5 \langle \partial_{\mu} U^{\dagger} \partial^{\mu}
U (U^+ M_0 + M_0^+ U)\rangle
+ L_6 \langle U^{\dagger} M_0 + M_0^+ U\rangle^2 \nn\\
+&& L_7 \langle U^{\dagger} M_0 - M_0^{\dagger} U \rangle^2 +
L_8 \langle M_0^{\dagger} U M_0^{\dagger} U + U^{\dagger}
M_0 U^{\dagger} M_0 \rangle ,
\nn
\end{eqnarray}}
where the terms which couple to external sources, like gauge
fields, are omitted \cite{chpt1,chpt2}. The $L_i$ constants are
related with the renormalized $L_i^r (\mu)$ generically as
$L_i=L_i^r(\mu)+\Gamma_i\lambda$ \cite{chpt2} where $\mu$ is the
renormalization scale,

\begin{equation}\lambda=\frac{ \mu^{d-4}}{16\pi^2} \left[
\frac{1}{d-4}-\frac{1}{2}\left(\log 4\pi - \gamma
+1\right)\right],\label{lambda}\end{equation} $\gamma$ is the Euler constant
 and the $\Gamma_i$ coefficients can be found
in \cite{chpt2}. We remark that the $L_3$ and $L_7$ constants are not
renormalized and are therefore scale independent,
 i.e, $\Gamma_3=\Gamma_7=0$.

Thus, up to fourth order one has to consider the tree level
diagrams from $O(p^2)$ and $O(p^4)$, together with the one-loop
diagrams in Fig.\ref{fig1:diagrams}. We stress that mass and wave
function renormalizations should be accounted for to the same
order. The latter are schematically represented by the tadpole
diagram (e) in Fig.\ref{fig1:diagrams}.
As customary, we define the bare fields in terms of the renormalized
ones as
 $\pi= Z^{1/2}_\pi \pi^{ren}$ and so on for the kaons and eta,
so that
scalar fields have finite canonical kinetic terms. Taking into account
all the different contributions from diagrams of type (e) in
Fig.\ref{fig1:diagrams} plus those tree level diagrams coming from
${\cal L}_4$, one obtains:
 \begin{eqnarray}
Z_\pi&=& 1+\frac{4}{3}\mu_\pi+\frac{2}{3}\mu_K
-\frac{4\lambda}{3 f_0^2}
\left(2M_{0\,\pi}^2+M_{0\,K}^2\right)\nonumber\\
&-&\frac{8}{f_0^2} \left[2 L_4^r M_{0\,K}^2+ (L_4^r+L_5^r)
M_{0\,\pi}^2\right],\nonumber\\
Z_K&=& 1+\frac{1}{2}\mu_\pi+\mu_K+\frac{1}{2}\mu_\eta
-\frac{2\lambda}{3 f_0^2}\left(M_{0\,\pi}^2+5M_{0\,K}^2\right) \nonumber\\
&-&\frac{8}{f_0^2} \left[(2 L_4^r+L_5^r) M_{0\,K}^2+ L_4^r
M_{0\,\pi}^2\right],\nonumber\\
Z_\eta&=& 1+2\mu_K-\frac{4\lambda}{f_0^2}M_{0\,K}^2 \nonumber\\
&-&\frac{8}{3
  f_0^2}\left[\left(3L_4^r-L_5^r\right)M_{0\,\pi}^2+
2\left(3L_4^r+2L_5^r\right)M_{0\,K}^2\right],
\label{zetas}
\end{eqnarray}
where
\be
\mu_i=\frac{M_i^2}{32\pi^2 f_0^2}\log\frac{M_i^2}{\mu^2},
\ee
with
$i=\pi,K,\eta$.

Note that the wave function renormalization constants
$Z_i$ contain a divergent part and they are scale dependent.

As for the mass renormalizations, the physical pion and kaon
masses are given in terms of the tree level ones as \cite{chpt2}:
 \begin{eqnarray}
M^2_\pi&=& M^2_{0\,\pi}\left[1+\mu_\pi-\frac{\mu_\eta}{3}+\frac{16
M_{0\,K}^2}{f_0^2}\left(2L_6^r-L_4^r\right)\right.\nonumber\\
&+&\left.
\frac{8
M_{0\,\pi}^2}{f_0^2}\left(2L_6^r+2L_8^r-L_4^r-L_5^r\right)
\right], \nonumber\\
M^2_K&=& M^2_{0\,K}\left[1+\frac{2\mu_\eta}{3}+\frac{8
M_{0\,\pi}^2}{f_0^2}\left(2L_6^r-L_4^r\right)\right.\nonumber\\
&+&\left.\frac{8
M_{0\,K}^2}{f_0^2}\left(4L_6^r+2L_8^r-2L_4^r-L_5^r\right)\right].
\nonumber\\
M^2_\eta&=& M^2_{0\,\eta} \left[1+2\mu_K-\frac{4}{3}\mu_\eta+
\frac{8M^2_{0\,\eta}}{f_0^2}(2L_8^r-L_5^r)\right.\nonumber\\
&+&\left.
\frac{8}{f_0^2}(2 M^2_{0\,K}+M^2_{0\,\pi})(2L_6^r-L_4^r)
\right]\nonumber\\
&+& M^2_{0\,\pi}\left[-\mu_\pi+\frac{2}{3}\mu_K+\frac{1}{3}\mu_\eta\right]
\nonumber\\
&+&\frac{128}{9f_0^2}(M^2_{0\,K}-M^2_{0\,\pi})^2(3L_7+L_8^r)
 \label{masses}
\end{eqnarray}

According to the chiral power counting, we have to use
Eq.(\ref{zetas}) and Eq.(\ref{masses}) only in the tree level part of
the amplitudes. In fact, the mass renormalization Eq.(\ref{masses})
affects only the mass terms coming from the Lagrangian
in Eq.(\ref{Lag2}) and not the masses coming from the kinematics of the
corresponding process. As it will be seen below, we will not need
the mass renormalization of $M_\eta$ in any of our expressions.

The meson decay constants are also modified to one loop. It will
be convenient for our purposes to write all the one-loop
amplitudes in terms of a single decay constant, which we have
chosen to be $f_\pi$. For that reason and for an easier comparison
with previous results in the literature, we also give here the
result for the meson decay constants to one loop \cite{chpt2}:
 \begin{eqnarray}
f_\pi&=& f_0\left[1-2\mu_\pi-\mu_K+\frac{4
M_{0\,\pi}^2}{f_0^2}\left(L_4^r+L_5^r\right)+\frac{8
M_{0\,K}^2}{f_0^2}L_4^r\right], \nonumber\\ f_K&=&
f_0\left[1-\frac{3\mu_\pi}{4}-\frac{3\mu_K}{2}-\frac{3\mu_\eta}{4}+\frac{4
M_{0\,\pi}^2}{f_0^2}L_4^r\right.\nonumber\\&+&\left.\frac{4
M_{0\,K}^2}{f_0^2}\left(2L_4^r+L_5^r\right)\right],\nonumber\\
f_\eta&=&f_0\left[1-3\mu_K+ \frac{4
L_4^r}{f_0^2}\left(M_{0\,\pi}^2+2M_{0\,K}^2\right)
+\frac{4M_{0\,\eta}^2}{f_0^2}L_5^r\right].
 \label{fpis}
\end{eqnarray}

It is important to stress that both the physical masses in
Eq.(\ref{masses}) and the decay constants in Eq.(\ref{fpis})
are finite and scale
independent.

Therefore, the one-loop ChPT scattering amplitude (renormalized and
scale independent)
for a given process will have the generic form:

\be
T(s,t,u)=T_2 (s,t,u)+T_4^{pol} (s,t,u)+T_4^{uni} (s,t,u) \nonumber
\ee where $s,t,u$ are the Mandelstam variables. Here, $T_2$ is the
tree level contribution from the Lagrangian in Eq.(\ref{Lag2}),
whereas
$T_4^{pol}$ contains the fourth-order terms which are polynomials
in $s,t,u$. Those polynomials have four possible origins: tree
level terms from the Lagrangian in  Eq.(\ref{lag4}) proportional  to
$L_i^r$, other polynomial terms 
proportional to $L_i$ with $i=4...8$ 
coming from the mass and decay constant renormalization in eqs.(\ref{masses})
and (\ref{fpis}), terms proportional to $\mu_i$ coming from tadpole
diagrams ((d) and (e) in Figure \ref{fig1:diagrams})
and finally pure
polynomial fourth-order terms which stem from our parametrization
of the one-loop functions (see Appendix A).
Let us remark that, for technical reasons explained 
in section III.B,
 we have chosen to write all our amplitudes in terms of 
$f_\pi$ only since, using eqs.({fpis}), $f_K$ and $f_\eta$ can be 
expressed in terms of $f_\pi$, $L^r_4$ and $L^r_5$.
In addition, $T_4^{uni}$ stands for the contribution of
diagrams (a),(b) and (c) in Figure \ref{fig1:diagrams}.
These contributions not only contain  the
imaginary parts required by unitarity but also yield the correct
analytic structure for the perturbative amplitudes, as it will be
discussed below. We remark that  all the terms in $T_4^{uni}$ will
be  proportional to the  $\bar J$ and $\jbb$ functions
defined in Appendix \ref{ap:form}.

Using crossing symmetry it is not difficult to see
that there are only eight independent meson-meson amplitudes.  We
have calculated these amplitudes to  one loop  in $SU(3)$ ChPT.
They are given in Appendix \ref{ap:amplitudes}. Three of these
amplitudes had not been calculated before, namely
$\bar{K^0}\eta\rightarrow \bar{K^0}\eta$,
$\eta\eta\rightarrow \eta\eta$ and $K\eta\rightarrow K\pi^0$.
For the rest, we have checked that our amplitudes
coincide with previous results \cite{Kpi,JAPaco} up to differences
in notation and different simplification schemes, equivalent
up to $O(p^6)$. In particular, since we are interested in the
``exact'' form of perturbative unitarity (see below),
we have written  our final results in terms of  a single pion decay
constant, $f_\pi$,  and we have
 used the Gell-Mann--Okubo relation taking care of preserving
exact perturbative unitarity.
Furthermore, we have explicitly checked that
all the amplitudes remain finite and scale independent.

Finally, we wish to add a remark about $\eta-\eta'$ mixing, since the physical
$\eta$ is indeed a mixture of the U(3) 
octet and singlet pseudoscalars, whereas in this work we are
only using the standard SU(3) ChPT. 
One may wonder then if our description of the $\eta$ is just that 
of the pseudoscalar octet component, since in this Lagrangian
the singlet field is not an explicit
degree of freedom. However, it has been shown \cite{LeutwylerNc} that
the standard framework  results from an expansion in powers of 
the inverse powers
of the ``topological susceptibility'' of the complete $U(3)$ Lagrangian.
In that context the $\eta'$ is considered as a massive state 
(that is why it does not count as an explicit degree of freedom)
but the singlet component generates a correction to $L_7$.
Note that, indeed, the mass of the $\eta$
contains an $L_7$ contribution, and that is why we can use $M_\eta$ in
eq.(\ref{masses}) with its physical value, whereas the $M_{0\eta}$ 
is the one satisfying the Gell-Mann-Okubo relation exactly.
Therefore, our approach can be understood as the lowest order approximation
to the $\eta-\eta'$ mixing problem, where all the effects of the mixing 
appear only through $L_7$.  Since we will only compare with data
in states with one $\eta$ at most, and below 1200 MeV, our results 
seem to suggest that this approximation, 
although somewhat crude, is enough with the present status of the
experimental data. Indeed, we will see that the
values that we obtain for $L_7$ are in perfect agreement
with those given in the literature (and this comparison can now be
done
because we have the complete one loop amplitudes 
renormalized in the standard way).

\section{Partial waves and unitarity}
\label{parwa}
\subsection{Partial waves}

Let us denote by $T^{IJ}_{ab} (s)$ the partial wave for the
process $a\rightarrow b$, i.e, the projection of the amplitude for
that process with given total isospin $I$ and angular momentum
$J$. That is, if $T^{I}_{ab} (s,t,u)$ is the isospin combination
with total isospin $I$, one has
\be
T^{IJ}_{ab} (s)=\frac{1}{32 N\pi}\int_{-1}^{1} d x P_J (x)
T^{I}_{ab} \left(s,t(s,x),u(s,x)\right)
\ee
where $t(s,x), u(s,x)$
are given by the kinematics of the process $a\rightarrow b$ with
$x=\cos\theta$,  the scattering angle in the center of mass frame.

Note that we are normalizing the partial waves including a factor
$N$, such that $N=2$ if all the particles in the process are
identical and $N=1$ otherwise. Recall that, since we are working
in the isospin limit, the three pions are considered as identical,
so that $N=2$ only for the  $\pi\pi\rightarrow \pi\pi$ and
$\eta\eta\rightarrow \eta\eta$ processes.

We shall comment now on the  $T^{I}_{ab}$ amplitudes for every
possible process involving $\pi,K,\eta$.  Using crossing
symmetry and assuming isospin symmetry
exactly, we will  determine the number of
independent amplitudes for each process. The discussion is general
and there is no need to invoke ChPT, although we will refer to the
results for the amplitudes in Appendix \ref{ap:amplitudes}, which
are the one-loop ChPT results.
\begin{itemize}

\item\noindent{\it $\pi\pi\rightarrow \pi\pi$ scattering:} There is only one
independent amplitude, so that one has

\begin{eqnarray}
T^0(s,t,u)&=&3\, T(s,t,u)+T(t,s,u)+T(u,t,s),\nonumber\\
T^1(s,t,u)&=& T(t,s,u)-T(u,t,s),\nonumber\\
T^2 (s,t,u)&=&T(t,s,u)+T(u,t,s),\nn
\end{eqnarray}
where $T(s,t,u)$ is the
$\pi^+\pi^-\rightarrow\pi^0\pi^0$ amplitude.
At one-loop in ChPT it is given in Appendix \ref{ap:amplitudes},
Eq.(\ref{amppipi}).

\item{\it $K\pi\rightarrow K\pi$ scattering:} Crossing symmetry
allows us to write the $I=1/2$ in terms of the $I=3/2$ one as
 \be
T^{1/2} (s,t,u)=\frac{1}{2} \left[3\,T^{3/2} (u,t,s)-T^{3/2}
(s,t,u)\right].
\ee

Here, $T^{3/2} (s,t,u)$ is the $K^+\pi^+\rightarrow K^+\pi^+$
amplitude, whose expression at one-loop within
 ChPT corresponds to Eq.(\ref{amppik}).

\item{\it $K\bar K\rightarrow K\bar K$ scattering:} We can write the
isospin amplitudes as
\begin{eqnarray}
T^0 (s,t,u)&=&T_{ch} (s,t,u)+T_{neu} (s,t,u), \nn\\
T^1 (s,t,u)&=&T_{ch} (s,t,u)-T_{neu} (s,t,u),
\end{eqnarray}
where $T_{ch}$ and $T_{neu}$ are, respectively, the amplitudes for
the processes  $K^+ K^-\rightarrow K^+ K^-$ and  $\bar K^0
K^0\rightarrow K^+ K^-$. Their expressions to one loop
correspond to Eqs.(\ref{amp4kch}) and (\ref{amp4kneu}), respectively.

\item{\it $K\bar K\rightarrow \pi\pi$ scattering:} In this case, one
has
 \begin{eqnarray}
T^0 (s,t,u)&=&\frac{\sqrt{3}}{2}\left[T^{3/2} (u,s,t)+ T^{3/2}
(t,s,u)\right],\nn\\
T^1 (s,t,u)&=&\frac{1}{\sqrt{2}}\left[T^{3/2}
(u,s,t)- T^{3/2} (t,s,u)\right],
\end{eqnarray}
where $T^{3/2} (s,t,u)$ is the $K^+\pi^+\rightarrow K^+\pi^+$
amplitude, given in Appendix \ref{ap:amplitudes} for
one-loop ChPT, Eq.(\ref{amppik}).

\item{\it $K \eta\rightarrow K\eta$ scattering:} This is a pure $I=1/2$
process. The one-loop amplitude can be read directly from
Eq.(\ref{ampetak})

\item{\it $\bar{K}K\rightarrow\eta\eta$ scattering:} 
This is an $I=0$ process that using crossing symmetry
can be obtained from the previous amplitude  as follows:
\be
T_{\bar{K}^0K^0\rightarrow\eta\eta}(s,t,u)=
T_{\bar{K}^0\eta\rightarrow\bar{K}^0\eta}(t,s,u).
\ee

\item{\it $K \eta\rightarrow K\pi$ scattering:} This is also an $I=1/2$
process, whose amplitude correctly normalized, is

\be
T^{1/2} (s,t,u)=-\sqrt{3}\, T_{\bar K^0 \eta\rightarrow \bar
K^0\pi^0} (s,t,u),
\ee
where the  one-loop expression for $\bar{K}^0 \eta\rightarrow \bar
K^0\pi^0$ can be found in Eq.(\ref{ampketakpi})

\item{\it $\bar K K \rightarrow \pi\eta$ scattering:}
This is a $I=1$ process related to the $\bar K^0\eta\rightarrow \bar
K^0\pi^0$  amplitude by crossing symmetry, i.e:

\be
T^1 (s,t,u)=-\sqrt{2}\, T_{\bar K^0\eta\rightarrow \bar
K^0\pi^0} (t,s,u).
\ee

\item{\it $\pi\eta\rightarrow \pi\eta$ scattering:} This is a pure
$I=1$ isospin amplitude whose one-loop ChPT expression can be read
directly from Eq.(\ref{amppieta}).

\item{\it $\pi\pi\rightarrow\eta\eta$ scattering} Now $I=0$
and the amplitude is obtained from the previous
one by crossing, as
\be
T_{\pi^0\pi^0\rightarrow\eta\eta}(s,t,u)=
T_{\pi^0\eta\rightarrow\pi^0\eta}(t,s,u).
\ee

\item{\it $\eta\eta\rightarrow \eta\eta$ scattering:} Here, $I=0$ and
the corresponding one-loop amplitude can also be read directly
from Eq.(\ref{amp4eta}).

\end{itemize}
In this paper we will be interested in the case when there are
several coupled states for a given choice of $I,J$, i.e:
the coupled channel case. In particular, with the above
normalization, the relationship between the $T$-matrix elements
$T^{IJ}_{ab}$ and the $S$-matrix ones is given for  two coupled
channels ($a,b=1,2$) by
 \begin{eqnarray}
S_{11}&=& 1+2i\,\sigma_1 T_{11}\label{sat11},\\
S_{22}&=& 1+2i\,\sigma_2 T_{22}\label{sat22},\\
S_{12}&=&S_{21}= 2i\,\sqrt{\sigma_1\sigma_2}\, T_{12}\label{sat12},
\end{eqnarray}
where  the $IJ$ superscripts have been suppressed to ease the
notation and  we have used that due to time reversal invariance,
$T_{ij}=T_{ji}$. Here, $\sigma_i=2q_i/\sqrt{s}$ where $q_i$ is the
center of mass momentum in the state $i$. Note that $\sigma_i$ is
nothing but the phase space of that state at $\sqrt{s}$. 
In the $I=0$ channel above the $\eta\eta$ threshold
we will use the corresponding generalization in the case of three channels.

\subsection{Unitarity}

The $S$ matrix should be unitary, i.e, $S S^\dagger=1$. In case
there is only one state available, that means that $S$ can be
parametrized in terms of a single observable, which is customarily
chosen as the phase shift.  For the case of two channels,  the
elements $S_{ij}$ are organized in a unitary $2\times 2$ matrix,
containing only three independent parameters. We will follow
the standard parametrization: 
\be
S=\left(
\begin{array}{cc} \eta
e^{2i\delta_1}&i\sqrt{1-\eta^2}e^{i\left(\delta_1+\delta_2\right)}\\
i\sqrt{1-\eta^2}e^{i\left(\delta_1+\delta_2\right)}& \eta
e^{2i\delta_2}\end{array}\right).
\ee where the $\delta_i$ are the
phase shifts and $\eta$ is the inelasticity.

The unitarity relation  translates into relations for the elements
of the $T$ matrix of a particularly simple form for the partial
waves. For instance, if there is only one possible state, "1", for
a given choice of $I,J$, the partial wave $T_{11}$ satisfies
Eq.(\ref{sat11}), so that unitarity means
\begin{equation}
\ima T_{11}  = \sigma_1 \,\vert \,T_{11}\; \vert ^2 \Rightarrow
\; \ima T_{11}^{-1} = - \sigma_1,
\label{uni1}
\end{equation}

 In principle, the
above equation {\it only holds above threshold} up to the energy
where another state, "2", is physically accessible. If there are
two states available, then the $T$ matrix elements satisfy
\begin{eqnarray}
\ima T_{11} &=& \sigma_1(s) \,\vert\, T_{11} \vert ^2+
\sigma_2 \,\vert\, T_{12} \vert ^2,\nonumber\\
\ima T_{12} &=& \sigma_1 \, T_{11}\,  T_{12}^*+
\sigma_2 \,T_{12}\, T_{22}^*,\nonumber\\
\ima T_{22} &=& \sigma_1 \,\vert\, T_{12} \vert ^2+
\sigma_2 \,\vert\, T_{22} \vert ^2. \nonumber
\end{eqnarray}
In matrix form they read,
\be
\ima T = T \, \Sigma \, T^*\; \Rightarrow\; \ima T^{-1} = -\Sigma,
\label{unimatrix}
\ee
with
\be
T=\left(
\begin{array}{cc}
T_{11}&T_{12}\\
T_{12}&T_{22} \\
\end{array}
\right)
\quad ,\quad
\Sigma=\left(
\begin{array}{cc}
\sigma_1&0\\
0 & \sigma_2\\
\end{array}
\right)\,,
\ee
which allows for an straightforward generalization to the
case of $n$ accessible states by using $n\times n$ matrices.

One must bear in mind that the unitarity relations imply that the
partial waves are bounded as the energy increases. For instance,
in the one channel case, from Eq.(\ref{uni1}) we can write:
\be
T_{11}=\frac{\sin\delta}{\sigma_1}e^{i\delta} \label{bound1}
\ee
where $\delta$ is the phase of $t_{11}$.

Note that all the unitarity relations, Eqs.(\ref{uni1}) and
(\ref{unimatrix}), are linear on the left hand side and
quadratic on the right. As a consequence, if one calculates the
amplitudes perturbatively as truncated series in powers of an
expansion parameter, say $T=T_2+T_4+\dots$, the unitarity
equations will never be satisfied exactly. In particular, for ChPT
that means that unitarity can only be satisfied {\it
perturbatively}, i.e.:
\begin{eqnarray}
\ima T_2 &=& 0,\nn\\
\ima T_4 &=& T_2 \,\Sigma\, T_2,
\label{pertuni}\\
...\nn
\end{eqnarray}
where the latter is only satisfied {\it exactly} if one is
careful to express $T_4$ in terms of masses and decay constants
consistently with the choice made for $T_2$.
That has not always been the case in the literature
and that is one of the reasons why we have recalculated some processes:
all our results satisfy exact perturbative unitarity.
Otherwise there are additional $O(p^6)$ terms in Eq.(\ref{pertuni}).
As we will see below, this 
will be relevant to obtain a simple formula for the unitarized
amplitudes. Our choice has been to rewrite all the $f_K$ and $f_\eta$
contained in the amplitudes in terms of $f_\pi$, $L^r_4$ and $L^r_5$
using the relations in eq.(\ref{fpis}).

The deviations from Eq.(\ref{unimatrix}) are more severe at high
energies, and in particular in the resonance region, since
unitarity implies that the partial waves are bounded, see
Eq.(\ref{bound1}), which cannot be satisfied by a  polynomial.
Generically, in the resonance region, the unitarity bounds are
saturated. If a polynomial is adjusted to saturate unitarity in a
given region, in general, it will break the unitarity bound right
afterward. Another way of putting it is that resonances are
associated to poles in the complex plane, that will never be
reproduced with polynomials.

For all these reasons, if we are interested in extending the good
properties of ChPT to higher energies, we have to modify the amplitudes,
imposing unitarity and a functional form
that allows for poles in the complex plane.
This will be achieved with the Inverse Amplitude Method.

\section{The coupled channel Inverse Amplitude Method}
\label{secIAM}

As it can be seen from
the unitarity condition in Eq.(\ref{unimatrix}), the imaginary
part  of the Inverse Amplitude is known exactly
above the corresponding thresholds, namely, $\ima T^{-1}=-\Sigma$.
Indeed, any amplitude satisfying the unitarity constraint should have
the following form:
\begin{equation}
T= \left(\rea T^{-1} - i \Sigma \right)^{-1}.
\label{preIAM}
\end{equation}
Consequently, we should only have to calculate the real part of $T^{-1}$.
As a matter of fact, many unitarization methods are just different
approximations to $\rea T^{-1}$ (see \cite{IAM2} for details).
The idea behind the Inverse Amplitude Method (IAM) is to use
the formula right above, but approximating $\rea T^{-1}$ with
ChPT. Since we have $T\simeq T_2+T_4+...$. Then
\begin{eqnarray}
  T^{-1}&\simeq& T_2^{-1}(1-T_4 T_2^{-1}+...),\\
\rea  T^{-1} &\simeq&  T_2^{-1}(1-(\rea T_4) T_2^{-1}+...),
\end{eqnarray}
so that  multiplying Eq.(\ref{preIAM}) by $T_2 T_2^{-1}$ on the
left and $T_2^{-1} T_2$ on the right, we find
\begin{equation}
T\simeq  T_2 (T_2-\rea T_4-i\, T_2\, \Sigma\, T_2)^{-1} T_2.
\end{equation}
At this point, if the amplitude satisfy ``exact perturbative
unitarity'', namely Eq.(\ref{pertuni}), we can simplify the above
equation to obtain a simple expression
\begin{equation}
 T\simeq T_2 (T_2-T_4)^{-1} T_2,
\label{IAM}
\end{equation}
This is the generalization of the IAM to coupled channels.
Note that this formula only ensures exact unitarity if
$T_4$ satisfies ``exact perturbative unitarity''.

The IAM was first applied to just one elastic channel \cite{Truong} and
it was able to reproduce  well the $\pi\pi$
and $\pi K$ scattering phase shifts below the $K \bar{K}$ and $K\eta$
thresholds, respectively. In addition it was able to generate the $\sigma$
(now called $f_0(400-1200)$) the $\rho$ and the $K^*$ resonances \cite{IAM1}.
Furthermore, it was shown how the formula for the one channel
IAM can be justified in terms of dispersion relations \cite{IAM1},
which allowed for the analytic continuation
to the complex plane and the
identification of the pole associated to each resonance
in the second Riemann sheet.

In view of Eq.(\ref{IAM}), it may seem necessary to know
the complete $O(p^4)$ ChPT calculation of each one of the $T$ matrix
elements. Nevertheless, one could use a further approximation
and calculate only the s-channel loops (Fig.1a), which are the only
responsible for the unitarity cut and are supposed to dominate
in the resonant region.  This was the approach  followed in \cite{IAM2},
having in mind  that  the complete ChPT calculations were
not available at that time for any meson-meson scattering two-channel
matrix. The results were
remarkable, reproducing  up to 1.2 GeV seven $(I,J)$
meson-meson scattering channels (17 amplitudes), and
even generating seven resonances.  However, the fact that the s-channel
loops were regularized with a cutoff, together with the omission of
crossed loops and tadpoles, made impossible
to compare the chiral parameters with those of standard ChPT
(still, they had the correct order of magnitude, as expected).
Besides, the low energy ChPT predictions were recovered only partially.
This motivated the authors in \cite{JAPaco} to calculate the
full $O(p^4)$ $K^+ K^-\rightarrow K^+ K^-$
and $K^+ K^-\rightarrow K^0 \bar{K}^0$ amplitudes\footnote{An erratum
for these amplitudes has appeared published when preparing this work.
The previous results and conclusions in \cite{JAPaco}
are nevertheless correct, since the errata did not affect the
numerical calculations. We thank J.A. Oller for discussions and
for letting us check that their corrected amplitudes
coincide with ours.}, which  allowed
for the unitarization  with Eq.(\ref{IAM})
of the $(I,J)$=$(0,0)$ and $(1,1)$ channels. This approach
yielded again a good high energy description but
also reproduced simultaneously
the low energy $\pi\pi$ scattering lengths. All these results
were obtained with $L_i$ parameters compatible with those of standard ChPT
\cite{JAPaco}.

As we have seen in the previous section, we have calculated
the last three independent $O(p^4)$ meson-meson scattering
amplitudes that were still missing. They are given in Appendix B
in an unified notation with the other five that we have recalculated
independently (correcting some minor misprints in the literature).
Therefore, we are
now ready to unitarize the complete meson-meson scattering by means
of Eq.(\ref{IAM}).

However, at this point  we have to recall that for a given energy,
Eq.(\ref{IAM}) has only been justified for a matrix whose dimension
is exactly the number of states accessible at
that energy. The reason is that the unitarity relation, Eq.(\ref{unimatrix}),
increases its dimensionality each time we cross a new threshold.
Thus, for instance, in $\pi\pi$ scattering, one should
use the one dimensional IAM up to the $K\bar{K}$ threshold,
then the two dimensional IAM, etc....although this procedure
yields discontinuities on each threshold, instead of a single
analytic function. Another possibility \cite{IAM2} is to use the IAM with the
highest possible dimensionality of the $I,J$ channel for all energies
\footnote{As a technical remark, let us note that in such case,
the IAM has to be rederived in terms
of the partial waves $T_{ab}$ divided by the CM momenta of
the initial and final states, to ensure that 
this new amplitudes are real at lowest order. From
there the derivation follows the same steps, and
we recover the very same Eq.(\ref{IAM}) by multiplying
by the initial and final state momenta in the end.}.
This second possibility yields an analytic (and hence continuous)
function, but
it may not satisfy unitarity exactly at all energies,
namely, when the number of opened channels is smaller than the
dimensionality of the IAM formula. Following with the $\pi\pi-K\bar{K}$
example, if we use the 2-dimensional IAM formula, we will have
exact 
unitarity ensured above the $K\bar{K}$ threshold, but not below.
In particular, if we still use the 2-dimensional
IAM below $K\bar{K}$ threshold, the IAM $\pi\pi$ scattering element
will have an additional spurious contribution from the imaginary part
of the $K\bar{K}$ scattering left cut, which extends up to
$\sqrt{s}=4(M_K^2-M_\pi^2)$.  This is a well known
and lasting problem in the literature \cite{Zinn,report,chinos}
that affects also other unitarization methods, like the K-matrix
\cite{report}. As a matter of fact, several years ago \cite{Zinn} it was
suggested that the physical solution  would probably be an
interpolation between
the two just mentioned approaches.
However, in the context of ChPT and the IAM, and for the $\pi\pi-K\bar{K}$
channels, it was found \cite{JAPaco}
that the violations of unitarity are, generically, of the
order of a few percent only. We have confirmed 
this result but now for the whole
meson-meson scattering sector. Even the threshold parameters can be accurately
reproduced, since they are defined through the real part of the amplitudes
which are almost not affected by the spurious part.
The origin of this problem is that the IAM in Eq.(\ref{IAM}) mixes the
left cuts of all the channels involved when performing the inverse of
the $T_2-T_4$ matrix. Thus, it  is not able to reproduce
the left-cut singularities correctly \cite{IAM1,PenningtonSa},
although numerically their contribution is negligible
when all the observables are expressed in terms of real parts of the
amplitude, and
{\it taking into account the present status of the data and the uncertainties
in the $L_i$}.

In this paper we have chosen to show the second approach,
since the one-dimensional IAM has been thoroughly studied in
\cite{IAM1}.
 Very recently there have been
proposed dispersive approaches \cite{chinos} to circumvent this problem
in the $\pi\pi$ $\bar KK$ system, but
they involve the calculation of left-cut integrals
that are hard to estimate theoretically.
It would be interesting to have them extended and related
to the ChPT formalism, but that is beyond the scope
of this work. The fact that we use the higher dimensional IAM formalism,
which contains spurious cuts, does not allow for a clean continuation
to the complex plane. Nevertheless, since poles
associated to resonances have already been found
in the one-dimensional case \cite{IAM1} and in other approximated coupled
channel IAM approaches \cite{IAM2}, we leave their description
for a generalized IAM approach with better analytic properties
\cite{inpreparation}. In this work we will concentrate on
physical $s$ values, and the compatibility of the unitarized description
of resonances and low energy data with existing determinations
of the chiral parameters. Nevertheless we will also show
that this can also be achieved with the first, discontinuous, approach.

\section{Meson-meson scattering data}
\label{secdata}

Let us then comment on the data available for each
channel:

{\bf Channel (I,J)=(1,1)} For the energies considered here,
the two states that may appear in this
channel are $\pi\pi$ and $K\bar K$. In Figs.2a and 3a, we plot the
data on the $\pi\pi$ scattering phase shift
obtained from \cite{Prot} and \cite{Esta}, which
correspond to the squares and triangles respectively.
Let us remark that the first set of data points tends to be between
two and three standard deviations higher than the second when the
phase shift is higher than 90 degrees, and the other way around for
smaller values of the phase shift (note that error bars are smaller
than the data symbols). Thus the data sets are not
quite consistent with one another, which could be fixed with the
addition of a systematic error of the order of a few percent.

This channel is completely dominated by the $\pi\pi$ state
and there is almost no inelasticity due to
$K\bar{K}$ production below 1200 MeV.
The  $(1-\eta_{11}^2)/4$ points from the inelasticity analysis
given in \cite{Ma} are shown in the lowest part 
of Figs.2.d and 3.d.

{\bf Channel (I,J)=(0,0)} For this channel we may have up to three
states, namely $\pi\pi$, $K\bar K$ and $\eta\eta$. In this case,
there are three observables with several sets of data, which, as
can be seen in Figs.2.b,c and d, are somewhat incompatible between
themselves when only considering the errors quoted in the
experiments. Again, they become compatible if  we assume a
systematic error of a few percent. For the $\pi\pi$ scattering
phase shift ($\delta_{00}$, see Fig.2.b), the experimental data
shown come from: different analysis of the CERN-Munich
Collaboration \cite{CERNmunich} (open square), as well as from
\cite{Prot} (solid square), \cite{Esta2} (solid triangle) and
\cite{Frog} (solid circle). Concerning the $\pi\pi\rightarrow
K\bar{K}$ phase shift, the data in Figs.2.c and 3.c corresponds to:
\cite{Ma} (solid triangle) and \cite{Cohen} (solid square) and
they are reasonably compatible, mainly due to the large errors in
the first set. Finally, we are also showing in Figs.2.d and 3.d the data for
$(1-\eta^2_{00})/4$, since it is customary to represent in that
way the values of the inelasticity $\eta_{00}$. The experimental
results are rather confusing here, mainly up to 1100 MeV, due to
problems in the normalization. 
From the data shown in the figure, we have only fitted to those
coming from: \cite{Cohen} (solid square),
\cite{Etkin} (solid triangle), \cite{polychronatos} (open square)
and \cite{wetzel} (open circle). There is a disagreement in the
normalization with the data of \cite{Lindebaum} up to a factor of
2 (see \cite{MoPe} for discussion). We have not included the latter
in the fit, mostly because in the analysis of \cite{Lindebaum} they
neglect the unitarity constraint, which in our approach is
satisfied exactly at those energies.

{\bf Channel (I,J)=(2,0)}
There is only the $\pi\pi$ state and so we only display
in Fig.2.e and 3.e
the $\delta_{20}$ phase shifts again
from the CERN-Munich Collaboration \cite{Hoogland} (open square)
 and the CERN-Saclay Collaboration \cite{Losty} (solid triangle).

{\bf Low energy $K_{l4}$ decay data.}
This reaction is particularly important since
it yields very precise information on the
$\delta_{00}-\delta_{11}$ combination of
$\pi\pi$ scattering phase shifts at very low energies.
In Figs.2.f and 3.f we show the data from
the Geneva-Saclay group \cite{Rosselet} (solid triangles)
and the very recent,
and more precise, data from E865 collaboration at Brookhaven
\cite{E865} (solid squares).

{\bf  Channel (I,J)=(1/2,1)} Here the possible states are
 $K\pi$ and $K\eta$.
We have plotted in Figs 2.g and 3.g data from the following experiments:
\cite{Mercer} (solid square) \cite{Esta3} (solid triangle).
Note that the first set is systematically lower than the
second, which is newer and more precise.
Nevertheless, they are compatible thanks mostly to the large error
bars on the first set.

{\bf  Channel (I,J)=(1/2,0)} Here the states are also $K\pi$ and $K\eta$.
The data in Figs 2.h and 3.h come from the following experiments:
\cite{Mercer} (solid square), \cite{Bingham} (open triangle), \cite{Baker}
(open diamond),
\cite{Esta3} (solid triangle), \cite{Aston} (open square).
It can be easily noticed that not all the data sets are compatible
within errors, but once again they can be reconciled by assuming a systematic
error of the order of a few percent.

{\bf Channel (I,J)=(3/2,0)} The only state here is $\pi K$.
In this case we have plotted in Figs.2.i and 3.i, data sets
from \cite{Esta3} (solid triangle) and \cite{Linglin} (solid square).
The latter is somewhat lower than  the former, although
they are compatible mostly due to the large errors in  \cite{Esta3}

{\bf  Channel (I,J)=(1,0)} The possible states for this case are
$\pi\eta$ and $K\bar K$.
We have plotted in Figs 2.j and 3.j the
$\pi\eta$ effective mass distribution  from the
$pp\rightarrow p(\eta\pi^+\pi^-)p$ reaction studied
by the WA76 Collaboration \cite{WA76}.
In order to reproduce this data, we are using
\begin{equation}
\frac{d\sigma_{\pi\eta}}{dE_{cm}}=c\, p_{\pi\eta} \vert T^{10}_{12}\vert^2+
\mbox{background}.
\label{c}
\end{equation}
Where the $c$ factor accounts for the normalization of the
mass distribution and
the dashed curve in those figures corresponds to a background due to
other resonances apart from the $a_0(980)$ (see \cite{WA76} for details).

\vspace{.5cm}

Once we have described the data on the different channels, we
will first compare with the IAM ``predictions'' from the present values of the
ChPT low energy constants, and later we will fit these
data by means of the IAM.

\section{The IAM with present low energy constant determinations}
\label{secIAM1}

In this section we will comment on the results of applying the
coupled channel IAM using the low energy constants from standard ChPT.
Since the values of these constants have been
determined from low energy data or large $N_c$ arguments,
the high energy results could
be considered as predictions of the IAM.
For our calculations we have used: $f_\pi=92.4\,$MeV, $M_\pi=139.57\,$MeV,
$M_K=495.7\,$MeV and
$M_\eta=547.45\,$MeV.

 In the second column of Table I we list the values obtained from
a very recent  and precise two-loop $O(p^6)$
analysis of $K_{l4}$ decays \cite{BijnensKl4}.
Note that the errors are only statistical. In the next column
we list the central values of the same analysis but only at $O(p^4)$.
In the fourth column we list the values from another set
where $L_1,L_2,L_3$ are taken from
an overall fit to $K_{e4}$ and $\pi\pi$ data \cite{BijnensGasser}
and  the rest are taken from \cite{chpt1}. Note that all of them
are pretty compatible and, except for $L_5$, the size of the error
bars is comparable.

\begin{table}[h]
\begin{tabular}{|c|c|c|c|}
&$K_{l4}$ decays $O(p^6)$&$K_{l4}$ decays $O(p^4)$& ChPT
\\ \hline
$L_1^r(M_\rho)$
& $0.53\pm0.25$
& $0.46$
& $0.4\pm0.3$ \\
$L_2^r(M_\rho)$
& $0.71\pm0.27$
& $1.49$
& $1.35\pm0.3$ \\
$L_3$
& $-2.72\pm1.12$
& $-3.18$
& $-3.5\pm1.1$ \\
$L_4^r(M_\rho)$
& 0
& 0
& $-0.3\pm0.5$ \\
$L_5^r(M_\rho)$
& $0.91\pm0.15$
& $1.46$
& $1.4\pm0.5$ \\
$L_6^r(M_\rho)$
& 0
& 0
& $-0.2\pm0.3$ \\
$L_7$
& $-0.32\pm0.15$
& $-0.49$
& $-0.4\pm0.2$ \\
$L_8^r(M_\rho)$
& $0.62\pm0.2$
& $1.00$
& $0.9\pm0.3$ \\
\end{tabular}
\caption{Different sets of chiral parameters $\times10^{3}$. The
second and third columns come from an $O(p^6)$ and  $O(p^4)$
analysis of $K_{l4}$ decays [42],
respectively.
Note that $L^r_4$ and $L^4_6$ are set to zero.
In the third column $L^r_1,L^r_2,L_3$ are taken from  [43]
and  the rest from  [2]
($L^r_4$ and $L^r_6$ are estimated from the Zweig rule). }
\label{eleschpt}
\end{table}

\nopagebreak
In Fig.2 we show the results of the IAM with the values given in
the fourth column of Table \ref{eleschpt}.
The solid curve corresponds to the central
values, whereas the shaded areas cover the uncertainty
due to the error on the parameters. They have been obtained
with a Monte-Carlo Gaussian sampling of 1000 choices of low energy constants
for each $\sqrt{s}$, assuming the errors are uncorrelated.
It is worth noticing that these error bands are so wide that
the results for the other columns in Table \ref{eleschpt} are rather similar,
even for the central values. Qualitatively all of them look the same.

It can be noticed that the IAM results, even with the
low energy parameters from standard ChPT,
already provide distinct resonant shapes of the
$\rho$,  $f_0(980)$, $K^*$, and $a_0(980)$
(see Figs. 2.a, 2.b, 2.g, and 2.j, respectively).
In addition, the IAM also provides two other extremely
wide structures
in the $(0,0)$ $\pi\pi$ and $(1/2,0)$ $\pi K$
scattering amplitudes. They correspond to
the  $\sigma$ (or $f_0(400-1200)$) and $\kappa$ (see Figs. 2.b and 2.h).
These structures are too wide to be considered as Breit-Wigner resonances,
but they are responsible for the relatively high values of the phase shifts
(the strength of the interaction) already near threshold.
In the last years there has been a considerable discussion about the 
existence and properties of these two states (for 
references, see the  scalar meson review in the PDG \cite{PDG}).
Since ChPT does not deal directly with quarks and gluons, it is very 
difficult
to make any conclusive statement about the spectroscopic nature 
of these states (whether they are $q\bar{q}$, four quarks states, 
meson molecules, etc...) unless we make additional assumptions 
\cite{nuestroNPA},
which would then spoil much of the model independency of our
approach, which is based just on chiral symmetry and unitarity.
Nevertheless, the simplicity and remarkable results of this
method gives a strong support, from the theoretical side, for the existence
of both the $\sigma$ and the $\kappa$. From previous works,
it is known that the ChPT amplitudes
unitarized with the IAM  generate the poles in the second Riemann sheet
associated with the $\sigma$ and the $\kappa$ 
 around $\sqrt{s_{pole}}\simeq 440-i 225\,\hbox{MeV}$ \cite{IAM1,IAM2} 
and 
$\sqrt{s_{pole}}\simeq 770-i 250\,\hbox{MeV}$ \cite{IAM2}, respectively.
(Let us remember that since these states are very wide, the familiar
relations $M\simeq\rea \sqrt{s_{pole}}$ and 
$\Gamma\simeq -2\ima\sqrt{s_{pole}}$ are very crude approximations).
We have checked that similar results are obtained for the amplitudes 
of this work. These values have to be considered as estimates,
since the uncertainties must be rather big, taking into account that the 
data in this channels are very conflictive (see Figs. 2 and 3).
The fact that we are able to reproduce these states with  parameters
compatible with previous determinations is also a strong
support for their masses and widths, which are in agreement
with recent experimental determinations both for the $\sigma$
and the $\kappa$ \cite{charm}.

To summarize, we have just shown how the present status
of both the experimental data and the $L_i$ determinations
allows for a use of the IAM despite the approximations made 
on its derivation, like the poor 
description of the left cut commented above.

\section{IAM fit to the scattering data}
 \label{secIAMfit}

Once we have seen that the IAM already describes the basic
features of meson-meson scattering, we can proceed to fit
 the data in order to obtain a more accurate description.
For that purpose we have used the MINUIT function minimization
and error analysis routine from the CERN program library \cite{MINUIT}.

Our results are presented in Fig.3, whose different curves and bands
can be understood as follows: As we have already seen
when commenting the experiments in the previous section, and as it
can be noticed in Figs.2 and 3, there are
several incompatible sets of data for some channels.
In the literature, this is usually solved by adding an extra systematic
error until these values are compatible. We have made
three fits by adding a $1\%$, $3\%$ and $5\%$ errors to the data on
each channel.  The continuous line corresponds to the $3\%$ case
and the resulting $L_i$  values are listed in the second column of
Table \ref{elesfit}. The shaded areas have been obtained again
from a Monte-Carlo sampling using the
$L_i$ uncertainties given by MINUIT for this fit,
which are listed on the third column of Table \ref{elesfit}.
Let us remark that there would be almost no difference to the
naked eye if we showed the fit with a $1\%$ or a $5\%$, neither
in the central continuous line nor in the shaded bands.
Furthermore, the
 $\chi^2/d.o.f$ for any of these fits is always $O(1)$.

However, although the curves may almost remain unchanged
when fitting with a different global systematic error, the
values of the $L_i$ come out somewhat different from each fit.
This is an additional source of error on the $L_i$ parameters,
listed on the fourth column of Table II. It can be noticed
that it dominates the uncertainty on the $L_i$.
For illustration, the area between the
dotted lines in Fig.3 corresponds to a Gaussian sampling
of the chiral parameters with the two sources of error added
in quadrature.

By comparing the $L^r_i$ from the IAM fit in Table II
with those of previous ChPT determinations (in Table I), we see that there
is a perfect agreement between them. This comparison of the
complete IAM fit parameters is only possible now that
we have the full $O(p^4)$ amplitudes, given in Appendix B,
which are regularized and renormalized following the same
scheme as standard ChPT. In particular, the 
agreement in the value of $L_7$  indicates that we are 
including the effects of the $\eta'$ consistently.


\begin{table}[h]
\begin{tabular}{|c|c||c|c|}
&Fit+errors& MINUIT error & From data \\
&(curve in Fig.3)& (band in Fig.3) & systematic error\\
\hline
$L_1^r(M_\rho)$
& $0.56\pm0.10$&$\pm 0.008$
& $\pm0.10$
\\
$L_2^r(M_\rho)$
& $1.21\pm 0.10$&$\pm 0.001$
& $\pm0.10$
\\
$L_3$
& $-2.79\pm0.14$ & $\pm0.02$
& $\pm0.12$
\\
$L_4^r(M_\rho)$ & $-0.36\pm0.17$
& $\pm0.02$
& $\pm0.17$
\\
$L_5^r(M_\rho)$
& $1.4\pm0.5$ & $\pm0.02$
& $\pm0.5$
\\
$L_6^r(M_\rho)$
& $0.07\pm0.08$ & $\pm0.03$
&  $\pm0.08$
\\
$L_7$
& $-0.44\pm0.15$&$\pm0.003$
& $\pm0.15$
\\
$L_8^r(M_\rho)$
& $0.78\pm0.18$& $\pm0.02$
& $\pm0.18$
\\
\end{tabular}
\caption{Low energy constants obtained from an IAM fit to the
meson-meson scattering data. The errors listed in the second column
are obtained by adding in quadrature those of columns 3 and 4.}
\label{elesfit}
\end{table}

The threshold parameters (scattering lengths and slope parameters)
 obtained with the IAM are given in Table
\ref{thresholdpar} for the low energy constants  in the second
column in Table \ref{elesfit}. The errors in Table \ref{thresholdpar}
are obtained by a Gaussian sampling of the above low energy
constants. Note that the experimental values
of the threshold parameters have not been used as input in the fit, and the
numbers we give are therefore predictions of the IAM.
As we have anticipated before and Table
\ref{thresholdpar} chows clearly, we are
able to reproduce the low energy behavior with great accuracy.
Let us then comment, for each different channel,
on the results of the IAM fit:

{\bf Channel (I,J)=(1,1)}
The most striking feature of this channel
is the $\rho(770)$ resonance, which, as it can be seen in Fig.3a,
can be fitted with a great precision. This had already been achieved
at $O(p^4)$
both with the single \cite{IAM1}  and the coupled \cite{JAPaco}
channel formalisms. However, this is now achieved in a
simultaneous fit with all the other channels, but
since we are using the
complete $O(p^4)$ expressions
we have a good description of the high energy
data without spoiling the scattering lengths  listed
in Table \ref{thresholdpar}.

This channel depends
very strongly on $2L_1^r+L_3-L_2^r$, and this combination can thus
be fitted with great accuracy.
The mass and width from a clear
Breit-Wigner resonance can be obtained from the phase shift by means of
\begin{equation}
\delta_{IJ}(M_R)=90^0\quad,\quad \Gamma_R=\frac{1}{M_R}
\left(\frac{d\delta_{IJ}}{ds}\right)^{-1}_{s=M_R^2}.
\label{BWmg}
\end{equation}
For the (1,1) case we obtain $M_\rho=775.7^{+4.3}_{-3.3}$ MeV
 and $\Gamma_\rho=135.5^{+8.0}_{-9.0}$ MeV ,
in perfect agreement with the values given in the PDG
\cite{PDG}. The errors correspond to a Gaussian sampling with the
central values quoted in the second column of Table \ref{elesfit}
and the MINUIT errors of the fit.

Finally, and just for illustration, the inelasticity
prediction from the IAM is shown in Fig.3d.
Note that the data values are so small
and the claimed precision is so tiny that any other effect 
not considered in this work (like the 4$\pi$ intermediate state)
would yield a contribution 
beyond the precision we can expect to reach with the IAM. That
is why they have been excluded from the fit.

{\bf Channel (I,J)=(0,0)}
There are three independent observables in this channel with data.
Concerning the $\pi\pi$ scattering phase shift, plotted
in Fig.3b,  we can reproduce
two resonant  structures. First, there is the $\sigma$ (or $f_0(400-1200)$),
which corresponds to a broad bump
in the phase shift, that gets as high as $50^0$
not very far from threshold.
This is not a narrow Breit Wigner resonance.
Indeed it was shown in the IAM with just one channel \cite{IAM1},
that it is possible to find an associated pole in the second Riemann sheet,
quite far from the real axis.
Second, we can nicely reproduce the shape of the
$f_0(980)$
which corresponds to a narrow
Breit-Wigner resonance although over a background phase provided
by the $\sigma$, so that its mass and width cannot be read directly from
Eq.(\ref{BWmg}).

Once more, it can be seen that
the scattering lengths can also be reproduced simultaneously
with the high energy data.

The next observable is the $\pi\pi\rightarrow K\bar{K}$ phase shift,
Fig.3c,
which can also be fitted neatly. Since we have included the $\eta\eta$
intermediate state, the fit is somewhat better than with just two
channels above the two $\eta$ threshold, as it was suggested in
\cite{JAPaco}, but not as much as expected (this could be due
to our crude treatment of $\eta-\eta'$ mixing, that we commented at the 
end of section II).

Finally, in Fig.3d, we show the inelasticity in the $(0,0)$ channel.
These are the most controversial sets of data, since there is a strong
disagreement between several experiments (up to a factor of 2 in the
overall normalization), as we have mentioned  when commenting
the data on this observable.

{\bf Channel (I,J)=(2,0)}
We have plotted the results in Fig.3.e.
Since only the $\pi\pi$ state can have these quantum numbers,
 we are simply reproducing the
single channel IAM formalism that already gave a very good description of
this non-resonant channel \cite{IAM2}. Nevertheless, let us remark
that it is now fitted simultaneously with all the other channels,
and the value of the
scattering length obtained from our fit is compatible
with the experimental result and standard ChPT, see Table III.

In addition, once we have a description of this and the $(0,0)$ channel,
we can obtain the phase of the $\epsilon'$ parameter which measures
direct CP violation in $K\rightarrow \pi\pi$ decays \cite{CP}.
It is defined, in degrees, as follows:
\be
\phi(\epsilon')=90^{\mbox{o}}-(\delta_{00}-\delta_{20})_{s=M_K^2}.
\ee
Our result is $\phi(\epsilon')=38\pm0.3$, where the error is obtained
from a Gaussian sampling of the parameters listed in column 2 of Table II
with the MINUIT errors
on the third column. This is in very good agreement with the
experimentally observed value of
$\phi(\epsilon')= 43.5\pm7$. Standard ChPT \cite{CPChPT}
predicts $45\pm6$.

{\bf Low energy $K_{l4}$ decay data.}
There is no real improvement in the description of
these low energy data in Fig.3f compared to ChPT, since standard
ChPT is working very well at these energies.
However, these very
precise data at so low energies ensure that the parameters of
our fit cannot be too different from those of standard ChPT.
In addition, they are extremely important in the determination
of the scattering lengths, in particular, of the controversial $a_{00}$.

{\bf  Channel (I,J)=(1/2,1)}
As it happened in the $(1,1)$ channel with the $\rho$, this channel is
 dominated by the $K^*(892)$. This is a distinct
Breit-Wigner resonance that can be fitted very accurately with
the IAM, see Fig.3g. From Eq.(\ref{BWmg}) we find
$M_{K^*}=889 \pm 5$ MeV
and $\Gamma_{K^*}=46\pm13$ MeV, in fairly good
agreement with the PDG \cite{PDG}. The errors have been obtained in
the same way as for the $\rho$ resonance in the (1,1) channel.

{\bf  Channel (I,J)=(1/2,0)}
Due to the wide dispersion of experimental results,
our fit yields a wide error band for this channel, as
it can be seen in Fig.3h. Nevertheless, as it happened
in the $(0,0)$ channel, the phase shift is of the
order of $50^0$ not far for threshold, due to a wide
bump similar to the $\sigma$ in that channel.
Here such a broad structure has been identified by different
experimental and theoretical
analysis \cite{kappa,IAM2,Sannino,charm} as the $\kappa$ although
there is still a controversy about its existence and origin
\cite{Penningtonkappa}, as it also happened with the $\sigma$.
It is very similar to the $\sigma$,
and hence it cannot be interpreted as a Breit-Wigner
narrow resonance.

We also give in Table III the value for the scattering length of this channel,
in good agreement with the experimental data, which
nevertheless is not very well known.

{\bf Channel (I,J)=(3/2,0)} Since only $\pi K$ can have  these
quantum numbers, this is once more the IAM with a single channel,
which already provided a very good description \cite{IAM1}. We
show in Fig.3i the results  of the global fit for this channel, as
well as the corresponding scattering length in Table III.

{\bf  Channel (I,J)=(1,0)}
In our global fit, the data in this channel, see Fig.3.j,
are reproduced using Eq.(\ref{c}). The shape of the
$a_0(980)$ is neatly reproduced in the mass distribution.
In order to compare the value of the normalization constant
$c$ with experiment, we also show in Fig.4
the result of applying the IAM with
the parameters obtained from our fit to the experimental data
obtained from $K^-p\rightarrow\Sigma^+(1385)\pi\eta$
and $K^-p\rightarrow\Sigma^+(1385)K \bar{K}$ \cite{Flatte}.
These data have not been included in our fit since they
do not have error bars, but it can be seen that the IAM
provides a good description.
 Once again we are using a formula like
 Eq.(\ref{c}), but with a  constant different from that for Fig.3.j
and no background. Our result is $c=63\pm15 \mu\mbox{b}/\mbox{GeV}$,
to be compared with the values quoted in   \cite{Flatte}
where $c$ was taken from $73$ to $165\, \mu\mbox{b}/\mbox{GeV}$.

{\bf Channel (I,J)=(0,1)} Finally, we show in Fig.5 the results
for the modulus of the amplitude in the $(0,1)$ channel. In this
case, there is only one meson-meson scattering channel, namely
$K\bar{K}\rightarrow K\bar{K}$. Therefore, we can only apply the
single channel IAM, and in so doing we find a pole at
approximately 935 MeV in the real axis. The width of this
resonance is zero, since  within our approach it can only couple
to $K \bar{K}$ and its mass is below the two kaon threshold. One
is tempted to identify this resonance with the $\phi(1020)$ meson,
but in fact it can only be related to its octet part $\omega_8$.
The reason is that the singlet part $\omega_1$ is SU(3) symmetric
and it does not couple to two mesons since their spatial function
has to be antisymmetric. Consequently we can only associate the
resonance obtained with the IAM to the octet $\omega_8$ 
\cite{IAM2,phinuestra}. The
position of the pole seems consistent with an intermediate mass
between the $\phi(1020)$ and the $\omega(770)$. This state had
also been found when using the IAM with the incomplete chiral
amplitudes \cite{IAM2}, and it had been used later to study the
$\phi\rightarrow\pi\pi$ decay within a chiral unitary approach
\cite{phinuestra}. The fact that we find it here again confirms
that it is not an artifact of the approximations used in
\cite{IAM2}. In addition although the amplitudes used here are
complete up to $O(p^4)$ and the fit is rather different, it
appears almost at the same place, which supports the soundness of
the results in \cite{IAM2}.

\begin{table}[h]
\begin{tabular}{|c|c|c|c|c|}
\hline
&Experiment&IAM fit&ChPT&ChPT \\
&&& ${\cal O}(p^4)$&${\cal O}(p^6)$\\
\hline \hline
$a_{0\,0}$&0.26 $\pm$0.05&0.231$^{+0.003}_{-0.006}$&0.20&0.219$\pm$0.005\\
$b_{0\,0}$&0.25 $\pm$0.03&0.30$\pm$ 0.01&0.26&0.279$\pm$0.011\\
$a_{2\,0}$&-0.028$\pm$0.012&-0.0411$^{+0.0009}_{-0.001}$&-0.042&-0.042$\pm$0.01\\
$b_{2\,0}$&-0.082$\pm$0.008&-0.074$\pm$0.001&-0.070&-0.0756$\pm$0.0021\\
$a_{1\,1}$&0.038$\pm$0.002&0.0377$\pm$0.0007&0.037&0.0378$\pm$0.0021\\
$a_{1/2\,0}$&0.13...0.24&0.11$^{+0.06}_{-0.09}$&0.17&\\
$a_{3/2\,0}$&-0.13...-0.05&-0.049$^{+0.002}_{-0.003}$&-0.5&\\
$a_{1/2\,1}$&0.017...0.018&0.016$\pm$0.002&0.014&\\
$a_{1\,0}$&&0.15$^{+0.07}_{-0.11}$&0.0072&\\
\hline
\end{tabular}
\caption{ Scattering lengths $a_{IJ}$ and slope parameters
$b_{IJ}$ for different meson-meson scattering channels. The
experimental data come from  [10,55],
the one loop results from
[5,8,10]
and those at two loops from [42].
We are using the definitions and conventions given in those
references.
 Let us
remark that our one-loop IAM results are closer
to those of two-loop ChPT, although the IAM depends on
much less parameters than the $O(p^6)$ ChPT.}
\label{thresholdpar}
\end{table}

Finally, we have also added in Fig.3 a dashed line that corresponds
to the result with the central values of the parameters in
the second column of Table II, but where we have used the one-channel
IAM at energies where there is only one state available, the two-channel
IAM when there are two, etc... As we commented at the end of section
\ref{secIAM}, this approach ensures exact unitarity at all  energies,
but we can see that it generates a discontinuity at each threshold.
The results are compatible within the wider error bands with the
previous IAM fit (the space between dotted lines).
This was expected since, as we have already commented,
the difference between the two approaches is of the order
of a few percent, which is also the order of magnitude of
the systematic error added to the data for the fit. Of course, it
is possible to obtain also a fit with this method, as it was done in
\cite{IAM1} and the resulting parameters are still
compatible with those listed in Table III.

\section{Conclusions}
\label{secsummary}

In this work we have completed the calculation
of the lightest octet meson-meson scattering amplitudes within
Chiral Perturbation Theory (ChPT) at one loop. We have calculated
three new amplitudes, $\eta\eta\rightarrow\eta\eta$,
$K\eta\rightarrow K\eta$ and $K\eta\rightarrow K\pi$
but we have also recalculated the other
five independent amplitudes, checking and revising previous results.
The full expressions are  given in Appendix B in a unified notation,
using dimensional regularization
and the $\overline{MS}-1$ renormalization scheme, which is the usual
one
within ChPT. All the meson-meson scattering partial waves below 1200 MeV,
with definite isospin $I$ and angular momentum $J$, can
be expressed in terms of those eight amplitudes.

Since ChPT is a low
energy theory, the one loop amplitudes have to be unitarized
in order to reach energies as high as 1200 MeV (and in particular
the two kaon  threshold). For that purpose we have applied
the coupled channel Inverse Amplitude Method, which ensures
unitarity for coupled channels and it is also able
to generate resonances and their associated poles,
without introducing any additional parameter. In addition,
it respects the chiral expansion at low energies, in our
case up to $O(p^4)$. Thus, we have shown how it is possible to
describe simultaneously the data on the $(I,J)=(0,0)$,
$(1,1)$, $(2,0)$, $(1,0)$, $(1/2,0)$, $(1/2,1)$,
$(3/2,0)$ meson-meson channels below 1200 MeV,
which correspond to 20 different reactions.
We also describe seven resonant shapes, namely,
the $\sigma$, $\rho(770)$, $K^*(892)$, $\kappa$,
$f_0(980)$,  $a_0(980)$ and the octet $\phi$.

This description is achieved with values
for the low energy constants which are perfectly compatible
with previous determinations obtained using standard ChPT and
low energy data. This comparison is only possible as far
as we now have the complete $O(p^4)$ expression for all
the amplitudes in the standard ChPT scheme.
Indeed, with the present determinations of standard ChPT,
we already find the resonance shapes and we obtain
the most distinct features of each channel, although
with big uncertainties due to the present knowledge
of the chiral parameters.

Nevertheless, we have performed a fit of our unitarized
amplitudes to the meson-meson data and we have obtained
a very accurate description
not only of the resonance region, but also
of the low energy data, and in particular of the scattering lengths.
We have also paid particular attention to the
uncertainties an errors in our description, which
have been estimated with Monte-Carlo samplings
of the fitted chiral parameters within their
resulting error bars.

Summarizing, we have extended and completed  previous analysis of
these techniques in the meson sector so that we believe that our
present work could be useful for further phenomenological
applications.

\section*{Acknowledgments.}

We are very grateful to J.A. Oller for his comments,
clarifications and discussions over almost every issue addressed
in this paper. We also thank J.Nieves and E.Ruiz-Arriola for
providing and explaining to us the Monte-Carlo code used to
generate the error bands. In addition, we have profited from
interesting discussions with A. Dobado and E. Oset. We acknowledge
partial support from the Spanish CICYT projects AEN97-1693, and FPA2000-0956,
PB98-0782 and BFM2000-1326.

\appendix

\section{Useful formulae}
\label{ap:form}

Here we will give the main results and definitions of the different
functions coming from the one-loop ChPT calculation. We are
following the notation and conventions of \cite{chpt2}.

When calculating the ChPT amplitudes, the typical loop integrals
that appear are, on the one hand, the tadpole integral, i.e, the
Feynman boson propagator evaluated at $x=0$:

\be
\int\frac{d^d q}{(2\pi)^d}\frac{i}{q^2-M_i^2}=2 M_i^2\lambda
+\frac{M_i^2}{16\pi^2}\log\frac{M_i^2}{\mu^2} \label{deltai} \ee
where $\mu$ is the renormalization scale, $i=\pi,K,\eta$,
 and we have extracted its divergent part
 for $d\rightarrow 4$, with $\lambda$ given in
(\ref{lambda}). On the other hand, the  integrals coming from
diagrams
 (a),(b) and (c) in Figure \ref{fig1:diagrams} is:

 \be
J_{PQ} (p^2)=-i\int\frac{d^d q}{(2\pi)^d}
\frac{1}{[q^2-M_P^2][(q-p)^2-M_Q^2]}\label{jotapq} \ee where
$P,Q=\pi,K,\eta$ and whose
divergent contribution in dimensional regularization can be
separated as

\be J_{PQ} (s)= J_{PQ} (0)+{\bar J}_{PQ} (s)+{\cal O} (d-4)\ee
where
\begin{eqnarray}
J_{PQ} (0)&=&-2\lambda-\frac{1}{16\pi^2}\frac{1}{\Delta}
\left[M_P^2\log\frac{M_P^2}{\mu^2}-M_Q^2\log\frac{M_Q^2}{\mu^2}\right]
\nonumber\\
{\bar J}_{PQ} (s)&=&\frac{1}{32\pi^2}\left[
2+\left(\frac{\Delta}{s}-\frac{\Sigma}{\Delta}\right)\log\frac{M_Q^2}{M_P^2}
\right.\nonumber\\ &-&\left.\frac{\nu (s)
}{s}\log\frac{(s+\nu(s))^2-\Delta^2}{(s-\nu(s))^2-\Delta^2}\right]
\label{jbarrapq}
\end{eqnarray}
and \begin{eqnarray}  \Delta&=&M_P^2-M_Q^2\nonumber\\
\Sigma&=&M_P^2+M_Q^2\nonumber\\\nu^2(s)&=&
\left[s-\left(M_P+M_Q\right)^2\right]
\left[s-\left(M_P-M_Q\right)^2\right]\nonumber
\end{eqnarray}

For the case of a single mass $M_P=M_Q$, the above integrals  read

\begin{eqnarray}
J_{PP}(s)&=&-2\lambda-\frac{1}{16\pi^2}\left(1+\log\frac{M_P^2}{\mu^2}\right)+
{\bar  J}_{PP} (s)\nonumber\\ {\bar J}_{PP}
(s)&=&\frac{1}{16\pi^2}\left[2+\sigma(s)\log\frac{\sigma(s)-1}{\sigma(s)+1}
\right]
\end{eqnarray}
with

\be
\sigma(s)=\left(1-4 M_P^2/s\right)^{1/2} \ee

Note that the above integrals have the correct unitarity structure
in the right cut which extends on the real axis from
$s=(M_P+M_Q)^2$ to infinity. In fact, all the integrals appearing
to one loop in ChPT can be expressed in terms of the tadpole and
$\bar J$ integrals above \cite{chpt2}. However, it is customary to
express the results also in terms of:

\be
\jbb_{PQ}(s)\equiv {\bar J} (s)-s {\bar J}' (0)
\ee
where, from (\ref{jbarrapq}) one has,
\be
{\bar J}' (0)=\frac{1}{32\pi^2}
\left[\frac{\Sigma}{\Delta^2}+2\frac{M_P^2
M_Q^2}{\Delta^3}\log\frac{M_Q^2}{M_P^2}\right]\ee

From the above definitions it is easy to check that the functions
$\bar J (s)/s$ and $\jbb (s)/s^2$ have well-defined limits as
$s\rightarrow 0$.

\onecolumn
\newpage

\begin{figure}[h]
\begin{center}
\hspace*{-.3cm}
\hbox{\psfig{file=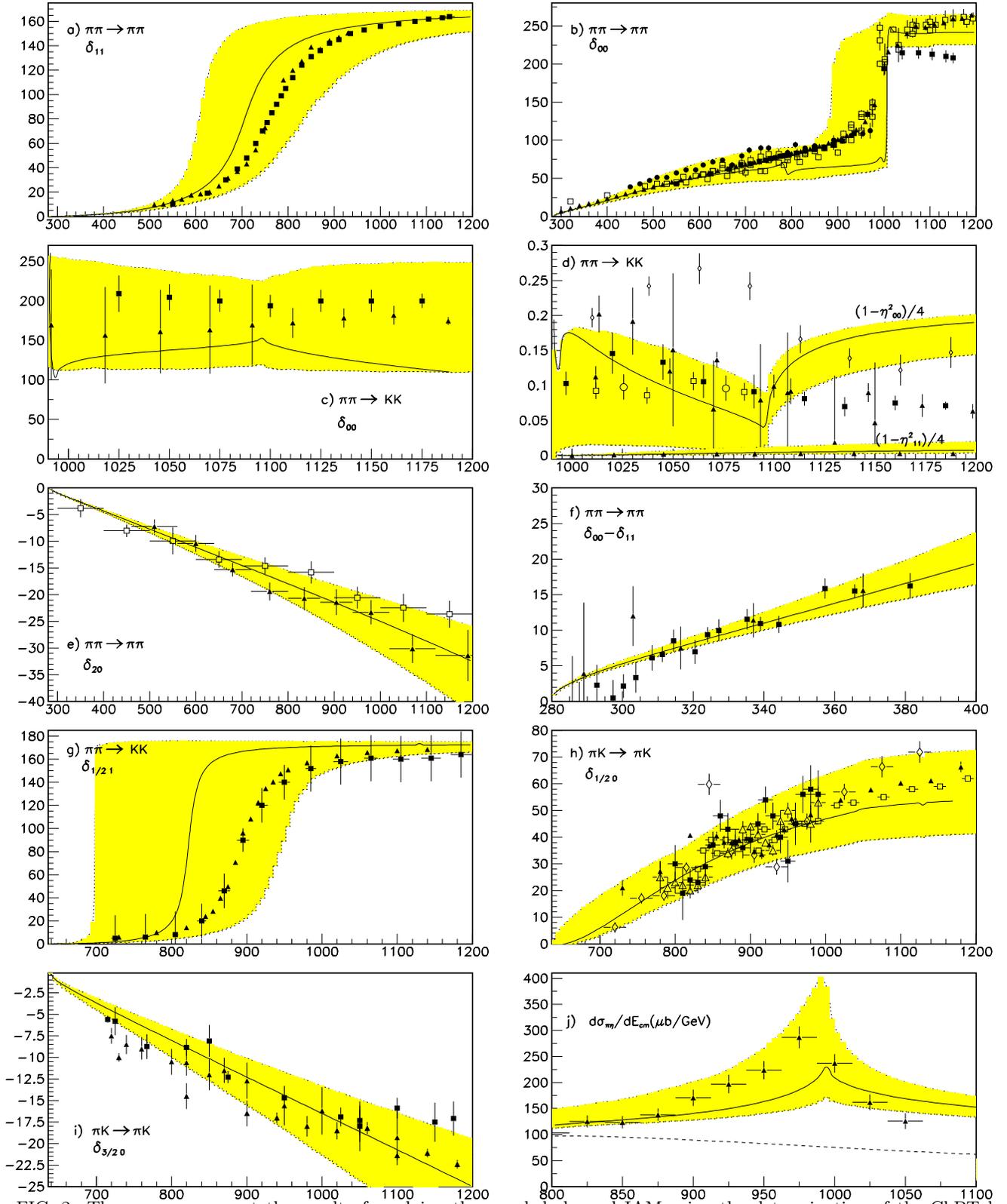,width=17.5cm}}
\caption{The curves represent the result of applying the coupled
channel IAM using the determination of the ChPT low energy
constant given in the fourth column of Table 1. The shaded area
covers the uncertainty due to the errors in those determinations
(assuming they were totally uncorrelated).} \label{fig2:figchpt}
\end{center}
\end{figure}

\newpage

\begin{figure}[h]
\begin{center}
\hspace*{-.5cm}
\hbox{\psfig{file=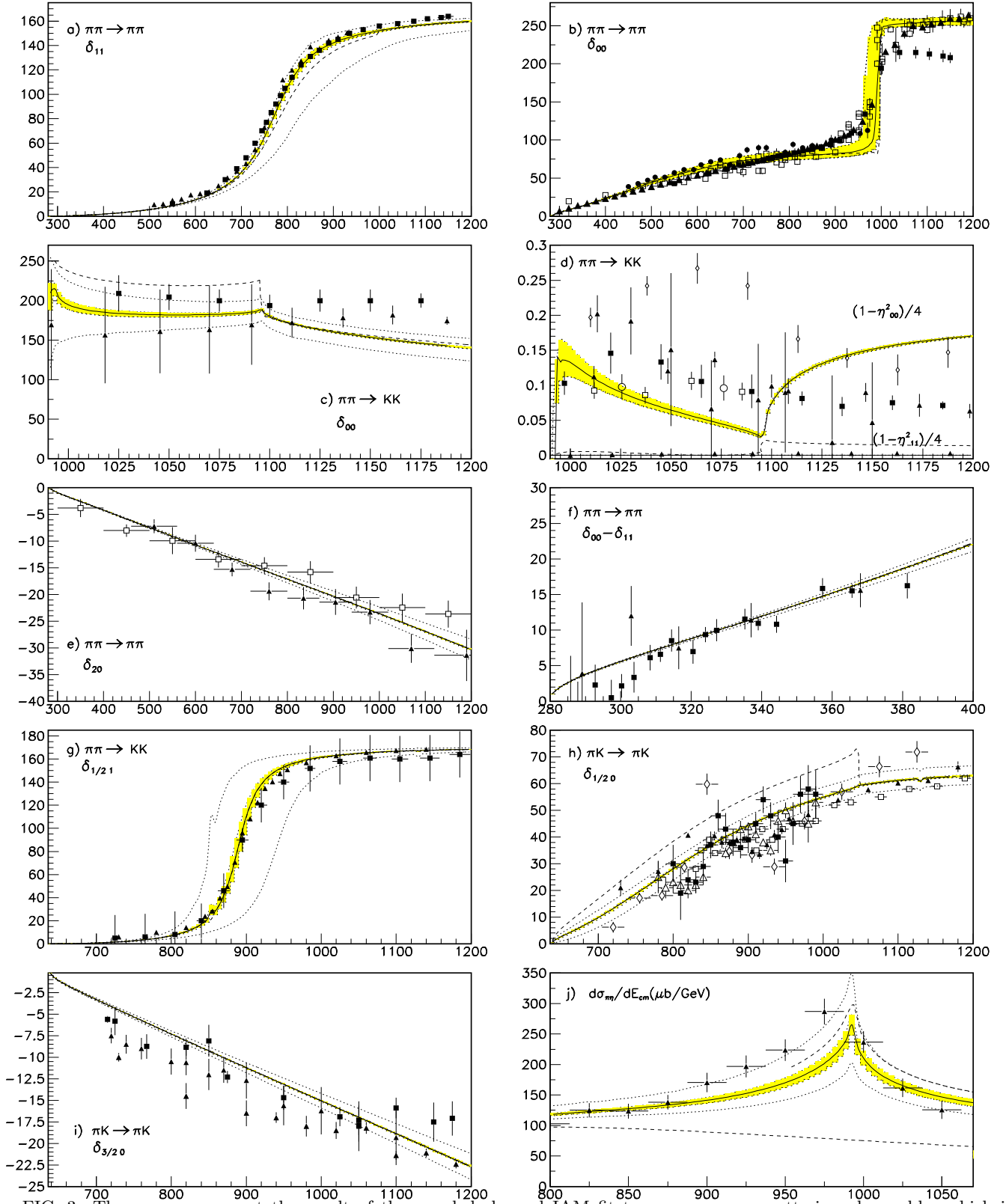,width=17.5cm}}
\caption{The curves represent the result of the coupled channel
IAM fit to meson-meson scattering observables which is described
in the text. The shaded area covers only the uncertainty due to
the statistical errors in the $L_i$ parameters obtained from
MINUIT (assuming they were uncorrelated). The area between dotted
lines corresponds to the error bands including in the $L_i$ the
systematic error added to the data (see text for details).
Finally, the dashed line corresponds to the use of the one channel
IAM when only one channel is accessible, but keeping the same
parameters of the previous fit.} \label{fig3:figfit}
\end{center}
\end{figure}

\newpage
\twocolumn

\begin{figure}
\begin{center}
\hspace*{-.3cm} \hbox{\psfig{file=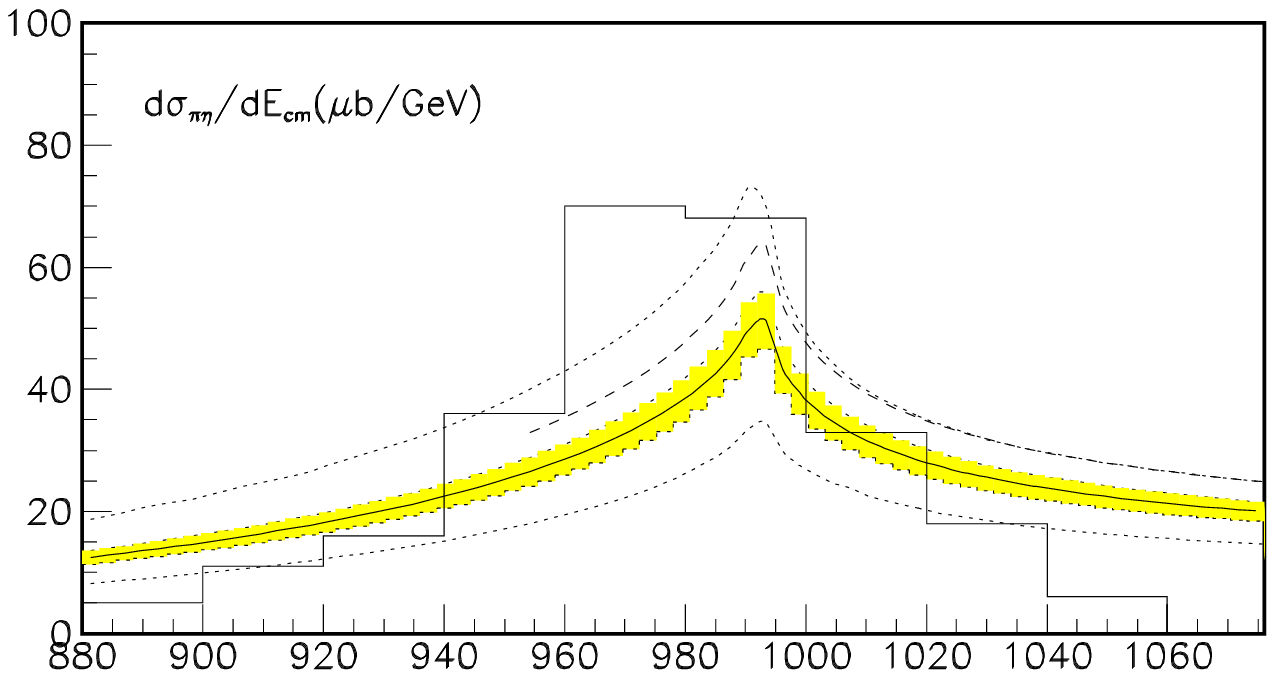,width=9.2cm}}
\hspace*{-.3cm} \hbox{\psfig{file=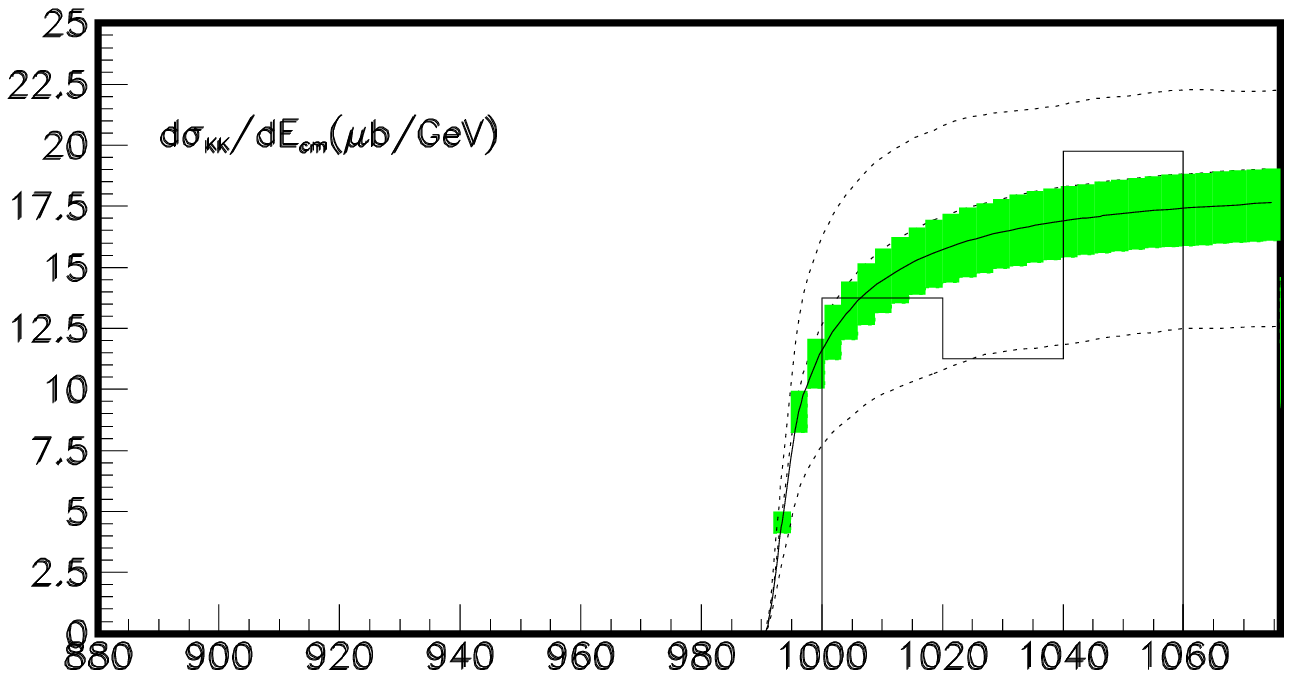,width=9.2cm}}
\vspace*{.1cm} 
\caption{ We show the effective mass distributions
of the two mesons in the final state of
$K^-p\rightarrow\Sigma^+(1385)\pi\eta$  (top) and
$K^-p\rightarrow\Sigma^+(1385)K \bar{K}$ (bottom), the data come
from  [53]. The curves and bands are as in Fig.3 }
\label{fig4:flatte}
\end{center}
\end{figure}

\begin{figure}
\begin{center}
\hbox{\psfig{file=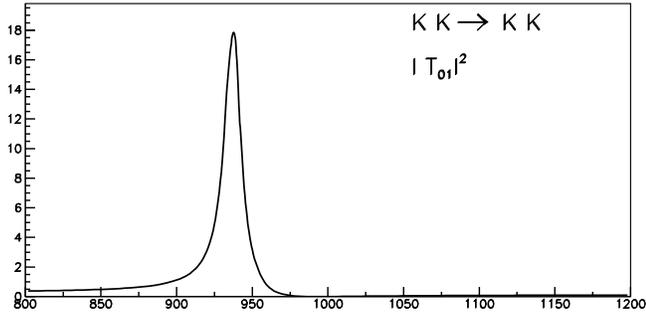,width=8.5cm}}
\vspace*{.1cm}
\caption{We show the modulus of the $(I,J)=(0,1)$
$K\bar{K}\rightarrow K\bar{K}$ amplitude.
The pole around 935 can be identified with the octet $\omega_8$
(see text for details).
Although that cannot be shown in a plot,
 the modulus of the amplitude actually becomes infinite.}
\label{fig5:phisolo}
\end{center}
\end{figure}

\newpage
\onecolumn
\section{One loop amplitudes from ChPT}
\label{ap:amplitudes}
\setlength{\arraycolsep}{.01cm}

Here we list the expressions of the eight independent meson-meson
scattering amplitudes to one loop in ChPT. We have carefully
checked the scale independence and perturbative exact unitary
(see section \ref{parwa}). Note that we have used eq.(\ref{fpis})
to write all the $f_K$ and $f_\eta$ in terms of $f_\pi$, $L^r_4$ and $L^r_5$,
in order to ensure ``exact'' perturbative unitarity, eq.(\ref{pertuni}).
Let us first
give the three amplitudes that had never appeared in the
literature in any form:

For $\eta\eta\rightarrow\eta\eta$:
\begin{eqnarray}
\label{amp4eta}
T(s,t,u)&=&\frac{16M_K^2-7M_\pi^2}{9f_\pi^2}+ \frac{\mu_\pi}{9
f_\pi^2}\left\{7M_\pi^2-48M_\eta^2\right\}
- \frac{\mu_K}{18 f_\pi^2M_K^2} \left\{81\left[t^2-s u-4 t
M_\eta^2 \right]+14M_\pi^4 -48M_\pi^2
  M_\eta^2\right.\\
&+&\left.378 M_\eta^4\right\} - \frac{\mu_\eta}{3 f_\pi^2
M_\eta^2}\left\{M_\pi^4-8M_\pi^2 M_\eta^2+ 24 M_\eta^4 \right\}
+\frac{4}{f_\pi^4} (2L_1^r+2L_2^r+L_3)\left\{s^2+t^2+u^2-4
M_\eta^4\right\}
\nonumber\\
&-&\frac{8}{3f_\pi^4}\left\{ 12 M_\eta^4
L_4^r+(3M_\pi^4-10M_\pi^2 M_\eta^2+13M_\eta^4)L_5^r -36 M_\eta^4
L_6^r -24(M_\pi^4-3M_\pi^2 M_\eta^2+2M_\eta^4)L_7 \right.
\nonumber\\ &-&\left.6 L_8^r (2M_\pi^4-6M_\pi^2M_\eta^2+7M_\eta^4)
\right\} -\frac{1}{192 \pi^2 f_\pi^4}\left\{27\left(t^2-s u-4 t
M_\eta^2\right)+ 16\left( 23 M_\eta^4 -22 M_K^2 M_\eta^2 +10
M_K^4\right) \right\}\nonumber\\
&+&\frac{1}{6f_\pi^4}\left\{ \frac{1}{27}(16M_K^2-7M_\pi^2)^2\bar
J_{\eta\eta}(s)+ M_\pi^4\bar J_{\pi\pi}(s) +\frac{1}{12}(9s-2
M_\pi^2-6M_\eta^2)^2\bar{J}_{KK}(s) + [s\leftrightarrow t] +
[s\leftrightarrow u] \right\}. \nonumber
\end{eqnarray}
For $\bar{K^0}\eta\rightarrow \bar{K^0}\eta$
\begin{eqnarray}
\label{ampetak}
T(s,t,u)&=&
\frac{9t-6M_\eta^2-2M_\pi^2}{12f_\pi^2} - \frac{2  L_5^r}{3
f_\pi^4}
       \left[ 3 M_\pi ^4 + 12 M_\eta^4 +
          M_\pi ^2\left( 5 M_\eta^2 -9t \right)  \right]
          \\
&+&\frac{1 }{3f_\pi^4}\left\{2\left( 12 L_1^r  + 5 L_3^r   \right)
\left( 2 M_K^2 -t \right)
     \left( 2 M_\eta^2 -t  \right)  +
    \left( 12 L_2^r +  L_3   \right)
     \left[ {\left( s -  M_K^2-  M_\eta^2 \right) }^2 +
       {\left( u-  M_K^2 -  M_\eta^2 \right) }^2
       \right]\right\}\nonumber\\
&+&\frac{4}{f_\pi^4}\left\{8 (L_6^r-L_4^r) M_K^2 M_\eta^2 +
    2  L_7\left(  M_\pi ^4 - 4 M_\pi ^2 M_\eta^2 +
       3 M_\eta^4 \right)  +  L_8^r
     \left(  M_\pi ^4 - 3 M_\pi ^2 M_\eta^2 + 6 M_\eta^4 \right)  +
   2  L_4^r t( M_\eta^2 +  M_K^2)
       \right\} \nonumber\\
&-&\frac{ \mu_\pi  }{48f_\pi^2
    \left(  M_K^2 -  M_\eta^2 \right) }\left\{2 M_K^2
         \left[26 M_\eta^2 + 69 t \right]  -84 M_K^4  +
        3\left[16 M_\eta^4- 50 t  M_\eta^2 +
           (s-u)^2 \right] \right\}\nonumber\\
&-&\frac{\mu_K  }{72f_\pi^2 M_K^2
    \left(  M_K^2 -  M_\eta^2 \right) } \left\{ 92 M_K^6 -
    81 M_\eta^2 t^2 -60 M_K^4
         \left[ 3 t  + M_\eta^2 \right] +
  18  M_K^2 (5 t^2 -2 s u +6 t M_\eta^2 + 8 M_\eta^4) \right\} \nonumber\\
&+&\frac{ \mu_\eta  }{144f_\pi^2 M_\eta^2
    \left(M_K^2 -  M_\eta^2 \right) } \left\{144 t M_K^4- 128 M_K^6
    +\left[27(s-u)^2-486 t M_K^2+428M_K^4 \right] M_\eta^2 \right.\nonumber \\
&+&\left.
       2\left[153\, t -166 M_K^2\right] M_\eta^4+144 M_\eta^6\right\}
+\frac{1}{2304f_\pi^4{\pi }^2}\left\{
   116 M_K^4+M_K^2\left[184 M_\eta^2-153\, t\right]
   \right.\nonumber\\
&-&\left.9 \left[10 t^2 + 2 s u -3 t
M_\eta^2+4M_\eta^4\right]\right\}
+\frac{t\bar{J}_{KK}(t)\left( 9t - 2 M_\pi^2 - 6 M_\eta^2 \right)
}{16f_\pi^4}
+\frac{ \bar{J}_{\eta\eta}(t)\left( 9t - 2 M_\pi^2-6M_\eta^2
\right)
      \left( 16 M_K^2 - 7 M_\pi^2 \right)  }{216f_\pi^4} \nonumber\\
&+&\frac{t\bar{J}_{\pi\pi}(t) M_\pi ^2}{8f_\pi^4}
+\frac{1}{32f_\pi^4} \left\{
\frac{\bar{J}_{K \eta}(s) }{9}\left[
27s(s-u)+189M_K^4+8M_\pi^4+54uM_\eta^2+
45M_\eta^4+12M_\pi^2(3s-2M_\eta^2)\right.\right.\nonumber\\
&-&\left.18M_K^2\left(6s-3u+4M_\pi^2+9M_\eta^2\right) \right]
+\frac{ \bar{J}_{K \pi}(s) }{9}\left[
27s(s-u)+29M_K^4+11M_\pi^4+18M_\eta^4\right.\nonumber\\&+&\left.
2M_K^2\left(18s+27u-47M_\pi^2-78M_\eta^2\right)+6M_\pi^2\left(9u-6s+8M_\eta^2\right)
  \right]
- \frac{ \bar{J}_{K \pi}(s)}{s}
\left[M_K^4\left(3u+14M_\pi^2-8M_\eta^2\right)\right.\nonumber\\
&+&\left.2M_K^6-2M_K^2
    M_\pi^2\left(3u+5M_\pi^2+4M_\eta^2\right)
+M_\pi^2\left(6M_\eta^4+M_\pi^2\left(3u+4M_\eta^2\right)\right)
\right] \nonumber\\
&+&\frac{\bar{J}_{K \eta}(s)}{s}{ \left( M_K^2 -  M_\eta ^2
\right) }^2
       \left( 4M_\pi^2 -18 M_K^2 -6 M_\eta ^2 -3u\right)
    \nonumber\\
&+&\left.6 \left(  M_K^2 - M_\eta ^2 \right)^2\frac{\jbb_{K
\pi}(s){\left(  M_K^2 - M_\pi ^2 \right) }^2  + \jbb_{K \eta}(s)
{\left( M_K^2 - M_\eta ^2 \right) }^2
    }{s^2} +[s\leftrightarrow u]\right\}. \nonumber
\end{eqnarray}

For $\bar{K}^0\eta\rightarrow \bar{K}^0\pi^0$:
\begin{eqnarray}
\label{ampketakpi}
T(s,t,u)&=&\frac{8M_K^2+3M_\eta^2+M_\pi^2-9t}{12\sqrt{3}f_\pi^2}+
\frac{\mu_\pi} {48\sqrt{3}f_\pi^2(M_K^2-M_\pi^2)} \left\{
27s^2+18su+27u^2+174tM_K^2\right.\\&-&\left.292M_K^4
+12(5M_K^2-6t)M_\pi^2-32M_\pi^4\right\}
-\frac{\mu_K}{24\sqrt{3}f_\pi^2 M_K^2(M_K^2-M_\pi^2)}\left\{
9t^2M_\pi^2+24M_K^6\right.\nonumber\\&+&\left.4M_K^4(17M_\pi^2-15t)
+2M_K^2\left[9(s-u)^2+6tM_\pi^2-22M_\pi^4\right]
\right\}-\frac{\mu_\eta}{16\sqrt{3}f_\pi^2 (M_K^2-M_\eta^2)}
\left\{3(s-u)^2\right.\nonumber\\
&+&\left.2(3t-14M_K^2+10M_\eta^2)(M_k^2-2M_\eta^2)\right\}
+\frac{1}{256\sqrt{3}\pi^2
  f_\pi^4}\left\{2(2s+u)(s+2u)-192M_K^4-23tM_\eta^2
-16M_\eta^4\right.\nonumber\\
&+&\left.5M_K^2\left(13t+24M_\eta^2\right) \right\}
-\frac{L_3}{\sqrt{3}f_\pi^4}\left\{s^2+4su+u^2-
30M_K^4-2tM_\eta^2+2M_\eta^4+\left.6M_K^2(t+2M_\eta^2)
\right\}\right.\nonumber\\
&+&\frac{1}{\sqrt{3}f_\pi^4}\left\{3M_\pi^4\left[L_5^r-2(2L_7+L_8^r)\right]
+M_\eta^4\left[6(2L_7+L_8^r)-L_5^r\right]-6L_5^rM_\pi^2(t-M_\eta^2)
\right\}\nonumber\\
&-&\frac{9t-8M_K^2-M_\pi^2-3M_\eta^2}{144\sqrt{3}f_\pi^4} \left[3
t \bar
 J_{KK}(t)+ 4 M_\pi^2\bar
J_{\pi\eta}(t)\right] +\frac{1}{288\sqrt{3}f_\pi^4}\left\{
  \bar J_{K\eta}(s)\left[27s(u-s)-45M_K^4+14M_\pi^4
\right.\right.\nonumber\\
&-&\left.\left.6M_\eta^2\left(9u+7M_\pi^2\right)-
9M_\eta^4+M_K^2\left(36s-54u+22M_\pi^2+156M_\eta^2\right)
   \right]+3 \bar
J_{K\pi}(s)\left[29M_K^4+7M_\pi^4\right.\right.\nonumber\\
&+&\left.\left.3s\left(9s+3u-4M_\eta^2\right)
 -
2M_K^2\left(16s+9u-18M_\pi^2+3M_\eta^2\right)
-M_\pi^2\left(40s+18u-30M_\eta^2\right)\right]\right.\nonumber\\
  &+&\left.9\frac{\bar J_{K\eta}(s)}{s}\left(M_K^2-M_\eta^2\right)
\left[10M_K^4+2M_\pi^4 -M_\eta^2\left(3u+8M_\pi^2\right)+
M_K^2\left(3u-12M_\pi^2+8M_\eta^2\right) \right]\right.\nonumber\\
&+&\left.9\frac{\bar J_{K\pi
}(s)}{s}\left(M_K^2-M_\pi^2\right)^2(3u-2M_K^2+2M_\pi^2)
- \frac{54
\jbb_{K\eta}(s)}{s^2}\left(M_K^2-M_\pi^2\right)\left(M_K^2-M_\eta^2\right)^3
\right.\nonumber\\ &-&\left. \frac{54\jbb_{K\pi
    }(s)}{s^2}\left(M_K^2-M_\pi^2\right)^3\left(M_K^2-M_\eta^2\right)
+[s\leftrightarrow u]\right\}. \nonumber
\end{eqnarray}

Apart from the above three amplitudes, we have recalculated the
other independent five. The reason is threefold. First, we wanted
them to satisfy exact perturbative unitarity to apply the simplest
IAM formulae. This was not the case of all the calculations in the
literature, even when considering the one-channel case. Second,
there have been several unfortunate misprints and errata in the
published formulae (including some errata made by one of the
authors). Finally we would like to have a self-contained
description of the one-loop calculation, together with all the
resulting formulae. Nevertheless, when compared with previous
analysis, our results are not exactly the same because we have
chosen to express the amplitudes in terms of only one physical
decay constant $f_\pi$, and we have only used the Gell-Mann--Okubo
relation to simplify masses if it did not affected the exact
perturbative unitarity relation. Apart from corrections, the differences
are $O(p^6)$.
The first amplitude to appear in the literature was
$\pi^+\pi^-\rightarrow\pi^0\pi^0$, although in SU(2) \cite{chpt1}.
However, we have been able to check also with the SU(3)
calculation \cite{Kpi}. The result, following the notation in
Appendix A, is:
\begin{eqnarray}
\label{amppipi}
T(s,t,u)&=&
\frac{s-M_\pi^2}{f_\pi^2}
-\frac{\mu_\pi}{3f_\pi^2M_\pi^2}\left\{4s^2-4tu-4 s M_\pi^2
+9M_\pi^4\right\}
-\frac{\mu_K}{6f_\pi^2M_K^2}\left\{s^2-t u+2 s M_\pi^2 \right\}
-\frac{\mu_\eta M_\pi^4}{9f_\pi^2M_\eta^2}\label{T4pi}\\
&+&\frac{4}{f_\pi^4}\left\{(2L_1^r+L_3)(s-2M_\pi^2)^2 +
L_2^r[(t-2M_\pi^2)^2+(u-2M_\pi^2)^2]\right\} \nonumber\\
&+&\frac{8M_\pi^2}{f_\pi^4}\left\{(2L_4^r+L_5^r)s+2(2L_6^r+
L_8^r-2L_4^r-L_5^r)M_\pi^2\right\} \nonumber\\
&+&\frac{1}{576\pi^2 f_\pi^4}\left\{30(M_\pi^2-s)s+21 t
u-56M_\pi^4\right\}
 +\frac{1}{2f_\pi^4}\left\{\frac{s^2\bar
J_{KK}(s)}{4}+ \frac{M_\pi^4\bar J_{\eta\eta}(s)}{9}
+(s^2-M_\pi^4)\bar J_{\pi\pi}(s)\right\}\nonumber\\
&+&\frac{1}{6f_\pi^4} \left\{\frac{(t-4M_K^2)(2s+t-4M_\pi^2)\bar
J_{KK}(t)}{4}+ \left[t(t-u)-2M_\pi^2(t-2u+M_\pi^2)\right]\bar
J_{\pi\pi}(t) +\left[t\leftrightarrow
u\right]\rule[.5cm]{0cm}{.2mm}\right\}. \nonumber
\end{eqnarray}

The $K^+\pi^+\rightarrow K^+\pi^+$ one loop calculation was first
given in \cite{Kpi}. It was correct up to $O(p^4)$ but when
expressed in terms of physical constants it did not satisfy  exact
perturbative unitarity. One of the authors gave an expression
satisfying that relation \cite{IAM1}, but there was also a
typographical error in that reference. Our corrected result,
expressed just in terms of $f_\pi$ is:
\begin{eqnarray}
\label{amppik}
T^{3/2}(s,t,u)&=&\frac{ M_K^2 + M_\pi^2-s }{2f_\pi^2}+
\frac{2 }{f_\pi^4} \left\{ (4 L_1^r+L_3)\left( t - 2M_K^2 \right)
       \left( t - 2M_\pi^2 \right)  +
       \left(2L_2^r+{L_3}\right)
         {\left( u - M_K^2 - M_\pi^2 \right) }^2  \right.
\\
&+&\left.
      2 {L_2^r} {\left( s - M_K^2 - M_\pi^2 \right) }^2
+  4 {L_4^r}\left[ tM_\pi^2 +
       M_K^2\left( t - 4M_\pi^2 \right)  \right]-
    2 {L_5^r}M_\pi^2\left( s + M_K^2 - M_\pi^2 \right) \right.\nonumber\\
&+&\left. 8\left( 2 {L_6^r} +  {L_8^r} \right) M_K^2M_\pi^2
\right\}
+\frac{\mu_\pi  }{24f_\pi^2M_\pi^2
    \left( M_\pi^2-M_K^2 \right) }
\left\{ 2M_K^2\left( 7s + 5u - 12M_K^2 \right)
       \left( 2M_\pi^2-t \right) \right.\nonumber\\
&-&\left.       \left[ 26s^2 + 21su + 25u^2 -
         3M_K^2\left( s + 5u + 16M_K^2 \right)  \right] M_\pi^2 +
      \left( 85s + 53u - 78M_K^2 \right) M_\pi^4 - 66M_\pi^6
      \right\}\nonumber \\
&+&\frac{ {\mu_K}}{12f_\pi^2M_K^2\left(M_\pi^2 - M_K^2 \right) }
\left\{ 42M_K^6 -
        M_\pi^2\left( 5s + 4u - 9M_\pi^2 \right)
         \left(2M_K^2-t \right)  +
        4 M_K^4\left(12M_\pi^2 -13s - 8u  \right)  \right.\nonumber\\
&+&\left.
        M_K^2\left[ 11s^2 + 12su + 7u^2 -
           3M_\pi^2\left( s - u + 8M_\pi^2 \right)  \right] \right\}
+\frac{\mu_\eta }{72f_\pi^2
    M_\eta^2\left( M_\eta^2 -M_\pi^2  \right) }
\left\{ 41M_\pi^6-18\left( s + u \right) M_\pi^4
\right.\nonumber\\
&+&\left.
      \left[ 36\left( s - u \right) u + 9\left( s + 5u \right) M_\pi^2 -
         59 M_\pi^4 \right] M_\eta^2 -
      3\left[ 9\left( 5s + u \right)  - 43M_\pi^2 \right]
       M_\eta^4 + 81 M_\eta^6 \right\} \nonumber\\
&+&\frac{1}{1152f_\pi^4
    {\pi }^2}
\left\{3\left( s - 10t \right) t - 6su - 3u^2 - 27M_K^4 +
    M_K^2\left( 30s - 3t + 21u - 34M_\pi^2 \right)  -
    3M_\pi^2\left(  t-2s  - 3u + M_\pi^2 \right) \right\} \nonumber\\
&-&\frac{\bar{J}_{K \pi}(u)}{16
    f_\pi^4}
      \left\{ \left( s - 5u \right) u + 5M_K^4 -
        2\left( s - 2u \right) M_\pi^2 + 5M_\pi^4 -
        2M_K^2\left( s - 2u + 5M_\pi^2 \right)  \right\} \nonumber\\
&+&\frac{\bar{J}_{\pi\pi}(t) }{24f_\pi^4} \left\{ t\left( 5t  -2s
+ 2M_K^2 \right)  +
      \left( 8s + 3t - 8M_K^2 \right) M_\pi^2 - 8M_\pi^4 \right\}
-\frac{ \bar{J}_{\eta\eta}(t)M_\pi^2}{72f_\pi^4}
      \left[ 2M_\pi^2 + 6{M_\eta }^2  -9t \right]\nonumber\\
&+&\frac{\bar{J}_{KK}(t) }{24f_\pi^4} \left\{
      M_K^2\left( 4s + 3t - 4M_\pi^2 \right)  +
      t\left( 4t-s + M_\pi^2 \right)  -4M_K^4 \right\}
+\frac{\bar{J}_{K\eta}(u)
  }{432f_\pi^4}\left\{2M_K^2
\left(27s+18u-74M_\pi^2-51M_\eta^2\right)\right.\nonumber\\
&+&\left. 29M_K^4+38M_\pi^4-M_\pi^2\left(36u-48M_\eta^2\right)
-9\left(3(s-u)u-6sM_\eta^2+M_\eta^4\right) \right\}
+\frac{\bar{J}_{K \pi}(s)}{4f_\pi^4}{\left( M_K^2+ M_\pi^2-s
\right) }^2 \nonumber\\
&-&\frac{\bar{J}_{K\eta}(u) }{48f_\pi^4 u}
\left\{2M_K^6+10M_\pi^4M_\eta^2
+3sM_\eta^4+M_K^4\left(3s-4M_\pi^2+10M_\eta^2\right)
-2M_K^2\left(2M_\pi^4+\left(3s+4M_\pi^2\right)M_\eta^2+3M_\eta^4\right)
\right\} \nonumber\\ &-&{\left( M_K^2 - M_\pi^2 \right) }^2
    \frac{ \bar{J}_{K \pi}(u)\left( s - 2M_K^2 - 2M_\pi^2 \right)
        }{16f_\pi^4u}
+ \left( M_K^2 - M_\pi^2 \right)^2 \frac{ \left( M_K^2 - M_\eta^2
\right)^2 \jbb_{K\eta}(u) + \left( M_K^2 - M_\pi^2 \right)^2
\jbb_{K \pi }(u)}{8f_\pi^4u^2}. \nonumber
\end{eqnarray}

The one-loop $\pi^0\eta\rightarrow\pi^0\eta$ amplitude was calculated
in \cite{Kpi}. We give here the result expressed in
terms of physical quantities:
\begin{eqnarray}
T(s,t,u)&=&
\label{amppieta}
\frac{M_\pi^2}{3f_\pi^2}
-\frac{\left( 13  M_\pi^4 + 6 t  \left(M_\pi^2-M_\eta^2\right) - 9
M_\pi^2 M_\eta^2 \right)\mu_\pi  }{9 f_\pi^2 \left(  M_\pi^2 -
M_\eta^2 \right) }
+\frac{ \left(  M_\pi^6 -  M_\pi^4  M_\eta^2 +
        4  M_\pi^2  M_\eta^4 \right) \mu_\eta  }{9 f_\pi^2
     M_\eta^2 \left( M_\pi^2 -  M_\eta^2 \right)}
\\
&&-\frac{ \mu_K }{6 f_\pi^2 M_K^2}
 \left\{ 20  M_\pi^2 \left( t - 3  M_\eta^2 \right) -25  M_\pi^4  +
        3 \left[ 3 \left( s^2 + s u + u^2 \right)  + 8 t  M_\eta^2 -
           9  M_\eta^4 \right]  \right\}
\nonumber\\
&&+\frac{4}{f_\pi^4}\left\{
2\left(L_1^r+L_3/6\right)(t-2M_\pi^2)(t-2M_\eta^2)+
(L_2^r+L_3/3)\left[(s-M_\pi^2-M_\eta^2)^2+(u-M_\pi^2-M_\eta^2)^2\right]
\right\}\nonumber\\
&&+\frac{8}{f_\pi^4}\left\{\left[t(M_\pi^2+M_\eta^2)-4M_\pi^2M_\eta^2\right]L_4^r
+2(2L_6^r-L_5^r/3) M_\pi^2 M_\eta^2+4
L_7M_\pi^2(M_\pi^2-M_\eta^2)+2L_8^r M_\pi^4\right\} \nonumber\\
&&+\frac{1}{576 f_\pi^4 {\pi }^2}\left\{77  M_\pi^4 +  M_\pi^2
\left( 154 M_\eta^2  -72 t\right) - 9 \left[ 3 \left( s^2 + s u +
u^2 \right) + 8 t  M_\eta^2 -
  9  M_\eta^4 \right]\right\}\nonumber\\
&&+\frac{1}{6f_\pi^4}\left\{
\frac{M_\pi^2}{9}\bar{J}_{\eta\eta}(t)(16M_K^2-7M_\pi^2)+
\frac{t}{4}\bar{J}_{KK}(t)(9t-2M_\pi^2-6M_\eta^2) +
\bar{J}_{\pi\pi}(t)M_\pi^2\left(2t-M_\pi^2\right)\right\}
\nonumber\\
&&+\frac{1}{9f_\pi^4}\left\{ M_\pi^4\bar{J}_{\pi\eta}(s)
+\frac{1}{24}\bar{J}_{KK}(s)\left(9s-8
M_K^2-M_\pi^2-3M_\eta^2\right)^2
+\left[s\leftrightarrow u\right]\right\}.
\nonumber
\end{eqnarray}

Finally, the $KK$ scattering amplitudes were calculated in \cite{JAPaco}.
They were given in a rather different notation from the previous ones.
Our result is, for $K^+K^-\rightarrow K^+K^-$:
\begin{eqnarray}
\label{amp4kch}T_{ch}(s,t,u)&=&
\frac{2M_K^2-u}{f_\pi^2}
-\frac{ \mu_K  }{6 f_\pi^2  M_K^2} \left[ 5 \left( s^2 + s t + t^2
\right)  + 6 u^2 - 13 u M_K^2 -
        8  M_K^4 \right]
\\
&+&\frac{\mu_\pi}{2 f_\pi^2} \left\{ 5\left( u - 2  M_K^2 \right)-
\frac{ 11 s^2 + 8 s t + 11 t^2 + 8 u  M_K^2 - 32  M_K^4  } {24
M_\pi^2} + \frac{ 9 \left( s^2 + t^2 \right)  + 24 u  M_K^2 -
 64  M_K^4  }{16( M_K^2 -  M_\pi^2)}
\right\}  \nonumber\\
&+&\frac{\mu_\eta} {12 f_\pi^2}
 \left\{ 64  M_K^2 - 2  M_\pi^2 - 27 u
- \frac{81 \left( s^2 + t^2 \right) - 36 \left( s + t \right)
M_\pi^2 + 8  M_\pi^4}{12 M_\eta^2} +
      \frac{ 9 \left( s^2 + t^2 \right)  + 24 u   M_K^2 -
           64  M_K^4 }{ 2(M_\pi^2 -  M_\eta^2)} \right\}
\nonumber \\
&+&\frac{4}{f_\pi^4}\left\{2L_2^r(u-2M_K^2)^2+(2L_1^r+L_2^r+L_3)
\left[(s-2M_K^2)^2+(t-2M_K^2)^2\right]
-4L_4^r u M_K^2-2 L_5^r(u-2M_K^2)M_\pi^2\right.\nonumber\\
&-&\left.4\left[L_5^r-2(2L_6^r+L_8^r)\right]M_K^4\right\}
+\frac{186 s t -177 u^2+1032 u M_K^2 -1648 M_K^4}{2304 f_\pi^4
{\pi }^2}
+\frac{1}{2f_\pi^4}(u-2M_K^2)^2\bar{J}_{KK}(u)\nonumber\\
&+&\frac{1}{288f_\pi^4}\left\{ 60 \left[ s \left( 2 s + t \right)
+ 4 u  M_K^2 -
     8 M_K^4 \right]\bar{J}_{KK}(s)
+2(9 s -8 M_K^2-M_\pi^2-3M_\eta^2)^2 \frac{\bar{J}_{\pi \eta
}(s)}{3} \right.\nonumber\\
&+&\left.(9 s -2 M_\pi^2-6M_\eta^2)^2 \bar{J}_{\eta \eta }(s)
+  3 \left[ s \left( 11 s + 4 t -8 M_K^2\right)  - 8 \left( s + 2
t -4M_K^2\right)  M_\pi^2 \right] \bar{J}_{\pi \pi }(s)+
\left[s\leftrightarrow t\right]
\rule[.5cm]{0cm}{.2mm}\right\}.\nonumber
\end{eqnarray}
And for $ \bar{K^0}K^0\rightarrow K^+K^-$:
\begin{eqnarray}
\label{amp4kneu}
T_{neu}(s,t,u)&=&
\frac{2M_K^2-u}{2f_\pi^2}
-\frac{ \mu_K  }{12 f_\pi^2  M_K^2} \left\{ 5 s^2- s u +8
u^2-2M_K^2\left(s+16u\right)+36 M_K^4
        \right\} \\
&+&\frac{ \mu _\pi}{4 f_\pi^2} \left\{ 5 \left( u - 2  M_K^2
\right)-
      \frac{ 11 s^2 + 4 t^2 + 4 s \left( 2 t + u \right)  -
           8 \left( s + 2 t \right)   M_K^2  }{12 M_\pi^2} +
      \frac{24 \left( s - 2 t \right)  M_K^2-9 \left( s^2 - 2 t^2 \right) +
        16  M_K^4}{8( M_K^2 -  M_\pi^2)} \right\}
         \nonumber\\
&+&\frac{\mu _\eta}
           {12 f_\pi^2}\left\{   9 \left( s - u \right)  + 14  M_K^2 -
 M_\pi^2   -
      \frac{{\left( 9 s - 2  M_\pi^2 \right) }^2}{12
      M_\eta^2}
      +
      \frac{
           8  M_K^2 \left( 3s-6t+2 M_K^2  \right)
-9 \left( s^2 - 2 t^2 \right) }
              {2\left( M_\pi^2 -  M_\eta^2\right)} \right\}
             \nonumber\\
&+&\frac{2}{f_\pi^4}
\left\{(4L_1^r+L_3)(s-2M_K^2)^2+2L_2^r(u-2M_K^2)^2+(2L_2^r+L_3)
\left(t-2M_K^2\right)^2\right\}\nonumber\\
&+&\frac{4}{f_\pi^4}\left\{4L_4^r s M_K^2 -2 M_K^4\left[4
L_4^r+L_5^r-2(2L_6^r+L_8^r)
\right]-L_5^r(u-2M_K^2)M_\pi^2\right\}\nonumber\\
&-&\frac{3 \left( 31 s^2 + 4 s u + 16 u^2 \right)  -
    4  M_K^2 \left( 30 s + 57 u - 80  M_K^2 \right) }{2304 f_\pi^4 {\pi }^2}
\nonumber\\
&+&\frac{{\bar{J}_{KK}}(s)}{6 f_\pi^4}\left[ s \left( s - u
\right) + 4 M_K^2 \left( 2 M_K^2 -t \right)  \right]
-\frac{{\bar{J}_{\pi \eta}}(s)}{432 f_\pi^4}(9 s - 8
M_K^2-M_\pi^2-3M_\eta^2)^ 2
+\frac{{\bar{J}_{KK}}(u)}{4 f_\pi^4}(u-2M_K^2)^2
 \nonumber\\
&+&\frac{{\bar{J}_{\pi \pi }}(s)}{96 f_\pi^4} \left[ s \left( 7 s
- 4 t+ 8 M_K^2 \right)+ 8 \left( s + 2t - 4 M_K^2 \right)  M_\pi^2
\right]
+\frac{{\bar{J}_{\pi \pi }}(t)}{24 f_\pi^4}\left( 2 s + t - 4
M_K^2 \right)  \left( t - 4 M_\pi^2 \right) \nonumber \\
&+&\frac{{\bar{J}_{\pi \eta }}(t)}{216 f_\pi^4}(9t -8
M_K^2-M_\pi^2-3M_\eta^2)^2
+\frac{{\bar{J}_{KK}}(t)}{24f_\pi^4}\left[t(s+2t)+4 u
M_K^2-8M_K^4\right]
+\frac{{\bar{J}_{\eta \eta}}(s)}{288 f_\pi^4}(9 s - 2
M_\pi^2-6M_\eta^2)^2 .\nonumber
\end{eqnarray}

\twocolumn


\begin{thebibliography}{99}
\footnotesize
\bibitem{weinberg} S. Weinberg, Physica A96, (1979) 327.
\bibitem{chpt1}J. Gasser and H. Leutwyler, Ann. Phys. 158, (1984)
142.
\bibitem{chpt2} J. Gasser and H. Leutwyler, Nucl. Phys. B250,
(1985) 465,517,539.
\bibitem{books} H. Leutwyler ``Chiral Dynamics''.
Contribution to the Festschrift in honor of B.L. Ioffe.
hep-ph/0008124.
A. Dobado, A.G\'omez-Nicola, A. L. Maroto and J. R. Pel\'aez,
{\it Effective Lagrangians for the Standard Model},
Texts and Monographs in Physics. ed: Springer-Verlag,
 Berlin-Heidelberg-New.York (1997).
A. Pich, Rept.Prog.Phys.58 (1995),563-610.
U.G. Mei{\ss}ner, Rept.Prog.Phys.56 (1993),903-996.
\bibitem{explicitresonances}
V. Bernard, N. Kaiser and U.G. Mei{\ss}ner, \NP{B364} (1991), 283.
J.A. Oller, E. Oset, Phys.Rev.D60:074023,1999.
M.Jamin, J.A. Oller, A.Pich,  Nucl.Phys.B587 (2000), 331-362.
\bibitem{LS} J.A. Oller, and E. Oset \NP{A620} (1997), 438.
J. Nieves, E. Ruiz Arriola. Phys.Rev.D63 (2001) 076001;
\PL{B455} (1999) 30; \NP{A679} (2000) 57.

\bibitem{Truong} T. N. Truong, \PRL{661}, (1988) 2526 ;\PRL{67}, (1991) 2260;
A. Dobado, M.J.Herrero and T.N. Truong, \PL{B235}, (1990) 134;

\bibitem{IAM1}
A. Dobado and J.R. Pel\'aez, \PR{D47}, (1993) 4883; \PR{D56}, (1997) 3057.
\bibitem{IAM2}J. A. Oller, E. Oset and
J. R. Pel\'aez, Phys. Rev. Lett. 80, (1998)
3452; Phys. Rev. D59, (1999) 074001; Erratum-ibid. D60, (1999) 099906.

\bibitem{Kpi}
V. Bernard, N. Kaiser, U.G. Mei{\it ss}ner, \PR{D43} (1991), 2757;
\NP{B357} (1991), 129;  \PR{D44} (1991), 3698.

\bibitem{JAPaco}  F. Guerrero and J. A. Oller, Nucl. Phys. B537, (1999) 459.
Erratum-ibid.B602, (2001), 641. 

\bibitem{LeutwylerNc} H. Leutwyler, \PL{B374}, (1996) 163
and \NP{B64} (Proc. Suppl), (1998) 223.
R. Kaiser, diploma work, Bern University, 1997.

\bibitem{GMO} M. Gell-Mann, Caltech Report CTSL-20 (1961).
S. Okubo, Prog. Theor. Phys 27 (1962) 949.



\bibitem{Zinn} D. Iagolnitzer, J. Zinn-Justin and J.B. Zuber,
\NP{B60} 233 (1973)

\bibitem{report} A. M. Badalyan, L. P. Kok, M.I. Polikarpov and
Y.A. Simonov, Phys. Rept. 82, 31 (1982).

\bibitem{chinos} Z. Xiao and H. Zheng, hep-ph/0103042 and hep-ph/0107188.

\bibitem{PenningtonSa} M. Boglione, M.R. Pennington. Z.Phys.C75,113 (1997);
I. P. Cavalcante, J. Sa Borges, hep-ph/0101037

\bibitem{inpreparation} A. Dobado, A. G\'omez-Nicola and J.R.Pel\'aez,
in preparation.

\bibitem{Prot} S. D. Protopopescu {\em et al.}, Phys. Rev. D7, (1973) 1279.

\bibitem{Esta} P. Estabrooks and A.D.Martin, Nucl.Phys. B79, (1974) 301.

\bibitem{Ma} A. D. Martin and E. N. Ozmutlu, Nucl. Phys. B158, (1979) 520.

\bibitem{CERNmunich}G.Grayer {\em et al.}, in {\it Experimental
Meson Spectroscopy},edited by A. H. Rosenfeld and K. W. Lai, AIP
Conf. Proc. 8 (AIP, New York, 1972) p. 5;
 G.Grayer {\em et al.}, presented at the 16th
Int. Conf. on High-Energy Physics, Batavia, 1972, paper No.768;
B. Hyams et al.\NP{B64},134 (1973);
 W.Manner, presented at the 4th Int. Conf. on Experimental
Meson Spectroscopy, Boston, MA, USA, April 1974,CERN
preprint; G. Grayer {\em et al.}, Nucl. Phys. B75, (1974) 189.

\bibitem{Esta2} P. Estabrooks {\it et al.}, in {\it $\pi\pi$ Scattering''},
edited by D.K. Williams and V.Hagopian, AIP Conf.
 Proc. 13 (AIP, New York,1973), p. 37.

\bibitem{Frog} C. D. Frogatt and J.L.Petersen, Nucl. Phys. B129, (1977) 89.



\bibitem{Cohen} D. Cohen, Phys. Rev. D22, (1980) 2595.


\bibitem{Etkin} A. Etkin {\it et al.} \PR{D25}, 1786 (1982).
\bibitem{polychronatos} V. A. Polychronatos {et al.} \PR{D19},1317 (1979)
\bibitem{wetzel} W. Wetzel {\it et al.}, \NP{B115}, 208 (1976)

\bibitem{Lindebaum} S. J. Lindebaum and R.S. Longacre,\PL{B274}, 492 (1992).

\bibitem{MoPe} D. Morgan and M.R. Pennington, \PR{D48},1185 (1993).

\bibitem{Hoogland} W. Hoogland {\it et al.}, \NP{B126}, 109 (1977)

\bibitem{Losty}M. J. Losty {\it et al.}, \NP{B69}, 185 (1974).

\bibitem{Rosselet}  L. Rosselet {\em et al.}, Phys. Rev. {D15}, (1977) 574.

\bibitem{E865} P. Tru\"ol for the E865 Collab., hep-ex/0012012,
M. Zeller for the  E865 Collab. "$\pi\pi$ scattering from $K_{l4}$ decays''
Talk presented in Chiral Dynamics 2000.
(Available at: http://www.jlab.org/intralab/calendar/chiral/).

\bibitem{Mercer} R. Mercer {\em et al.}, \NP{B32}, (1971) 381.

\bibitem{Esta3} P. Estabrooks {\em et al.}, \NP{B133}, (1978) 490.

\bibitem{Bingham} H. H. Bingham {\em et al.}, Nucl. Phys. B41, (1972) 1.


\bibitem{Baker} S. L. Baker {\em et al.}, \NP{B99}, (1975) 211.

\bibitem{Aston} D. Aston {\em et al.}
Nucl. Phys. B296, (1988) 493.

\bibitem{Linglin} D. Linglin {\em et al.}, \NP{B57},64 (1973).

\bibitem{WA76} T.A. Armstrong  {\em et al.}, \ZP{C52}, 389-396 (1991)
\bibitem{BijnensKl4} G. Amor\'os, J. Bijnens and P. Talavera,
\NP{B585},293,(2000); Erratum-ibid.B598,665 (2001);
\NP{B602},87,(2001).

\bibitem{BijnensGasser} J. Bijnens, G. Colangelo and J. Gasser,
\NP{B427}, (1994) 427.

\bibitem{nuestroNPA} J.R. Pel\'aez, J.A. Oller and E. Oset, \NP{A675}
(2000) 92c.

\bibitem{PDG} The Particle Data Group, {\it Review of Particle Physics},
Eur. Phys. J. C15, 1-878 (2000).

\bibitem{charm} E791 Collaboration,\PRL{86},(2001) 770.
C. Gobel for the E791 Collab. hep-ex/0012009, and
talk given in IX International Conference on Hadron Spectroscopy, 
HADRON2001, Protvino, Russia, August 2001 (to appear in the proceedings).

\bibitem{MINUIT}  F. James, Minuit Reference Manual D506 (1994)

\bibitem{CP} T.T. Wu and C. N. Yang, \PRL{13},380 (1964)

\bibitem{CPChPT} J. Gasser and U.-G. Mei{\ss}ner, \PL{B258} 129 (1991).

\bibitem{kappa} R.L. Jaffe, \PR{D15} 267 (1977); \PR{D15}, 281 (1977).
E. van Beveren {\it et al.} \ZP{C30}, 615 (1986).
S. Ishida {\it et al}, Prog. Theor. Phys. 98,621 (1997).

\bibitem{Sannino} D. Black, A. H. Fariborz, F. Sannino, J. Schechter. 
\PR{D58}:054012,1998. M. Jamin, J.A.Oller, A. Pich \NP{B587},331 (2000) 
 E. van Beveren, G. Rupp, hep-ex/0106077.

\bibitem{Penningtonkappa} S.N. Cherry, M.R. Pennington.
Nucl.Phys. A688, 823 (2001).


\bibitem{Flatte} S. M. Flatt\'e, \PL{B63} 224 (1976).

\bibitem{phinuestra} J.A. Oller, E. Oset and J. R. Pel\'aez, \PR{D62},
114017 (2000).


\bibitem{scatlength} O. Dumbrajs {\it et al} \NP{B191} 301 (1981).
\end{thebibliography}
\end{document}